\documentclass[letterpaper]{article}
\usepackage[utf8]{inputenc}
\usepackage{url}
\usepackage{amssymb}
\usepackage{amsmath}

\usepackage{geometry}
\usepackage{graphicx}
\usepackage{caption}
\usepackage{floatrow}
\graphicspath{{Plots/}}
\usepackage[section]{placeins}
\usepackage{float}
\usepackage{wrapfig}
\usepackage{appendix}
\usepackage{enumerate}
\usepackage{enumitem}
\usepackage{pgfplots}
\pgfplotsset{width=0.8\paperwidth,height=8.5cm,compat=1.9}
\usepackage{tikz}

\begin{document}

\author{Peter M\"attig \\
            Universit\"at Bonn}
\title{Validation of the Simulation of Collision Events at the LHC}
\maketitle



\tableofcontents

\include{acronym}

%
%
%



\abstract{The procedures of validating simulation of particle physics events at the LHC are
summarized. Because of the strongly fluctuating particle content of LHC events and detector 
interactions, particle based Monte Carlo methods are an indispensable tool for data analysis. 
Simulation in particle physics is founded on factorization and thus its global validation can
be realized by validating each individual step in the simulation. This can be accomplished by 
adopting results of previous measurements, in - situ studies and models.
Important in particle physics is to quantify how well simulation is 
validated such that a systematic uncertainty can be assigned to a measurement.  
The simulation is tested for a wide range of processes and agrees with 
data within the assigned uncertainties.}

\section{What Particle Physics is about: Example LHC}
\label{sec:particlephys}

During the 1960's and 70's the then differing concepts of electromagnetic, weak and strong interactions  
were put on a common ground leading to a theoretical framework, the Standard Model (SM)~\cite{bib:SM}. 
It is based on Quantum Field Theory and is arguably the most encompassing and
precisely tested theory of nature ever. It accommodates all measurements in an energy range of several
 ~100 GeV~\footnote{Using quantum mechanical relations this energy range can be interpreted as 10$^{-18}$ m,
i.e. about 100 million times smaller than an atom} with just a few fundamental particles, which can be 
classified in three sectors 

\begin{itemize}
\item Twelve spin 1/2 particles (fermions) that can be separated in quarks and leptons are 
          interpreted as 'matter' particles.
\item  Three kinds of spin 1 particles (vector bosons), the photon ($\gamma $), the $W$,  $Z^0$, and gluons,
          that transmit the electromagnetic, weak and strong interactions of the 
          matter particles, respectively.
\item  One spin 0 particle (scalar boson), the Higgs boson, to generate the masses of the fermions, force carriers
          and  itself.                      
\end{itemize}

The SM predicts the complete dynamics of these particles based on 19 free parameters\footnote{The number
increases if the masses of the neutrinos are considered. Not all of the parameters related to neutrinos 
have been measured. Since they do not affect physics at the LHC, the subject of this paper, they will not
be considered further.}. These elements and concepts appear to be the basis for nucleons, atoms and eventually
all phenomena beyond and are crucial for our current understanding of cosmology.

\subsection{The status of the Standard Model}
\label{sec:SMstatus}

The development of the conceptual framework was followed by an intensive experimental and theoretical
program to establish the existence of all SM components, measure the free parameters~\cite{bib-PDG} 
and test their dynamical properties. All components of the SM have been observed,
the last one, the Higgs boson, has been found in 2012 at the Large Hadron Collider (LHC) \cite{bib:LHC}
of the European Center for Particle Physics (CERN, Geneva),
and almost all allowed interactions have been
confirmed. The free parameters are measured to very high precision, e.g. 
the values of the coupling strength of the strong interaction and the mass of the $Z^0$, which 
will be used later, are determined as  

\begin{equation}\label{eq:as-MZ}
\alpha_s(M_Z) \ = \ 0.1181\pm 0.0011 , \ \ \ M_Z \ = \ 91.1876\pm 0.0021 \ \ \mathrm{GeV}
\end{equation}

i.e. they are know to the 10$^{-2}$ and 10$^{-5}$ level, respectively. 
Furthermore, despite 
intensive searches, no effect has been observed at accelerators that does not agree with the SM.
Therefore physicists consider the SM as being confirmed as a theory for an energy scale of up to 
several 100 GeV. The SM is also internally consistent, i.e. void of any infinities even 
at energies many orders of magnitude beyond what can be probed in any foreseeable future.

Still, there is a reluctance to accept the SM as a final theory. 
For one, the number and pattern of its particles and parameters are unexplained.  
In addition the SM does not contain gravity, which, however, should only become relevant at
M$_{\mathrm{Planck}}$ = 10$^{19}$ GeV, and it is not able to explain astrophysical observations
like the existence of Dark Matter~\cite{bib:DM} and Dark Energy~\cite{bib:DE}.
To address these issues, models have been devised that lead to physics beyond the SM (BSM).
Searches for BSM physics have so far been futile, but they have moved more and more into the 
focus of experimental and theoretical activities.

\subsection{The forefront Experiment: LHC}
\label{sec:LHC}

Searches for BSM physics require higher precision on SM processes and in particular
higher energies.
The energy frontier of particle physics is the LHC, where every 25 nano - seconds a bunch of 10$^{11}$ protons crosses 
another bunch flying in the opposite direction. Currently each of the protons has an energy of 
6.5 TeV leading to the highest collision energies reached in an accelerator and a huge rate of collisions.
Protons are composed of subcomponents, quarks or gluons, denoted together as partons.
At the LHC the protons serve only as vehicles for the partons and high energetic
parton collisions are of the main interest.
They produce a large variety of very different physics processes.

Each LHC bunch crossing leads to 'an event' consisting of a spray of some 1000 particles, 
which are recorded by highly precise and sophisticated
detectors covering almost the whole solid space angle and organized in different components.
Their electronic recordings are translated into 'physics objects',
i.e. candidates for electrons, quarks, photons etc. - each one a stable SM particles.
The content of an event, i.e.
how many of these different objects are found and their relation, e.g. relative angle or mass, 
make up the 'event signature'. 

In this article the ATLAS experiment~\cite{bib:ATLAS} will be considered as the example\footnote{
The simulation for other LHC experiments CMS~\cite{bib:CMS} and the specialized experiments 
ALICE~\cite{bib:ALICE}
and LHCb~\cite{bib:LHCb} work along fairly similar principles.}. 
It extends 40 m along the 
beam line and has a diameter of 25m in the transverse plane. 
Each of its components is sensitive to a particular type of particle, offering redundant and rather
comprehensive information. The wide coverage of the particles from the interaction
allows ATLAS to probe almost all of the LHC physics.
In a sense ATLAS combines some 100 distinct experiments of previous times. 
This allows it to make optimal use of each event including cross calibrations, an essential point
for the validation of the simulation - as will be discussed later.

The large number and variety of particles in each event
reflect the statistical properties of quantum mechanics.
The main experimental challenge is
to infer the underlying parton scattering from the different event signatures.  
It is at this point that simulation as a tool for
data analysis enters the field. 
Describing the large number of particles, their complicated and statistically distributed structure, as well as 
the statistical nature of their interactions in the detector cannot be achieved analytically 
but requires numerical methods. 

The article will start with a short overview over the principles of data analysis and the use of
simulations in particle physics, it will then discuss the two basic ingredients that enter simulation,
the modeling of the physics process and that of the detector. It will then formalize 
the principles of simulation to motivate validation procedures in particle physics 
before discussing examples of validation in some detail  
and how simulation and validation works in typical analyses. 
Finally some points raised in the philosophical literature on simulations are commented.  

\section{Data Analysis and the Use of Simulations}
\label{sec:SimuLHC}

Before describing the validation process in more detail in Sect.~\ref{sec:valid}, we will summarize the
ingredients of the simulation in particle physics. The simulation is particle based,
involving a wide range of different scales and is realized with computer codes applying Monte Carlo (MC)
methods.

\subsection{From Data to Physics}
\label{sec:DatPhys}

The principal aim of the simulation of LHC events is to understand how the physics process ('signal', S) of
interest would look in the detector. Comparing the measurement with this expectation allows one
to extract information about the physics process of interest, e.g. a parameter of the SM,
dynamic properties, or evidence, respectively exclusion of BSM effects. 

Each $S$ has a certain event signature. However, such signature is not 
unique but can also be due to competing processes, the 'background B'. 
To reach high sensitivity analyses aim at a good S/B ratio by applying
special selections exploiting different properties of the S and the B processes.
Arriving at a physics conclusion from the data requires the measurement to be compared to theory 
and thus to simulate both $S$ and $B$. 

In this article basic concepts of validation will often be discussed along the production of a Higgs ($h$),
and the detection via $Z^0$ bosons and electrons\footnote{For simplicity particles and anti-particles will 
just be denoted by the name of the particle.}. 

\begin{equation}\label{eq:hdecay}
g + g \ \rightarrow  \ h \ \rightarrow \ Z^0Z^{0*} \ \rightarrow \ (e^+e^-)(e^+e^-) 
\end{equation}

i.e. two gluons $g$ produce a Higgs that decays into two $Z^0$ bosons\footnote{The notation $Z^{0*}$ means that the boson
is 'off shell', i.e. its mass is different from the default 91 GeV due to quantum mechanical uncertainty.}, each of which then
decays into pairs of electrons.
The Higgs signal competes with background processes that also lead to four electrons but 
have no relation to Higgs production.

A key observable
for the Higgs production and decay, but also of general importance, is the cross section $\sigma_S$,
essentially a measure of the production yield. 
The cross section of a signal process is obtained from the measurement by

\begin{equation} \label{eq:Xsec}
\sigma_S \ = \ \frac{N_{\mathrm{sel}}  \ - \ N_{\mathrm{B}} }{{\cal{L} } \cdot \epsilon_S }
\end{equation}

where ${\cal{L} }$ is the luminosity, the total number of all possible proton - proton 
collisions provided by the 
LHC, $N_{\mathrm{sel}}$ is the number of data events, i.e. with four electrons, 
that have been selected, $N_{\mathrm{B}}$, the number of background events
and $\epsilon $, the efficiency to detect a selected signal event, i.e.

\begin{equation}\label{eq:eff}
\epsilon_S \ = \ \frac{N_{\mathrm{sel} }^{\mathrm{S} } }  {N_{\mathrm{prod}}^{\mathrm{S} }  }
\end{equation}

where $N_{\mathrm{sel}}^{\mathrm{S}}$, $N_{\mathrm{prod}}^{\mathrm{S}} $ are the numbers
of signal events that are selected, respectively, produced. 

\subsection{The Role of Simulation for Data Analysis}
\label{sec:roleSim}

Both $N_B$ and $\epsilon$ are obtained from simulation and should be determined 
to a high precision.
Per year some 10 trillion events are simulated for ATLAS, each taking 
several minutes. To make this work 100000 CPUs and 100s of petabytes
of disk storage are provided by a world wide computing grid. 
To guarantee a constant quality, the simulation is continuously checked with
bench mark processes.  

Simulations in particle physics are used for several purposes.

\begin{itemize}
\item In the data analysis they are instrumental to obtain model predictions and their
         experimental signatures that can be 
         compared to measurements. 
\end{itemize}

This is the broadest and most challenging application that will be primarily considered in this article.
In addition simulation is used to optimize tools and strategies.

\begin{itemize}          
\item Both in experimental and theoretical studies simulation is used to evaluate the feasibility of a 
         technique.
\item Simulations are used to optimize the lay - out of future detectors, or detector conditions to run 
         an experiment.                
\end{itemize}

In all three use cases simulation is an auxiliary method to experiments, 
but in no way replaces measurements with real experiments. 
E.g. even if a detector component is first devised with simulation, it is only after 
studying an actual prototype, that the component will be integrated into the experiment.
Simulations allow complicated manipulations of fairly involved models. 
Their results are believable only to the extent
that the input and structure of simulations is believable, which in turn requires
the input to and the structure of simulation to agree with experimental data. I.e.
validation of the simulation is essential.
 
In particle physics, two main parts have to be simulated, the underlying physics processes 
leading to a distribution of hadrons, photons and leptons, which can be measured in the detector, 
and the detector response to each of these particles. 
Both parts will be introduced in turn in the subsequent sections.

\section{Modeling the LHC processes}
\label{sec:phys-mod}

A general LHC collision of two partons $A,~ B$ producing a signal $S$, which
decays into two  particles $ C,~ D$, can be written as

\begin{equation} \label{eq:ABCD}
A+B \ \rightarrow \ S \ \rightarrow \ C+D
\end{equation}

A characterizing parameter for the parton collisions is the hardness of a collision
$Q^2$, which is given generically by $Q^2 = (p_C - p_A)^2$.
Here $p_i \ = \ (E_i,\textbf{p}_i)$ are the four momenta of the particles, with $E$ the energy and 
$\textbf{p}_i$ the momentum components in the three space directions.
'Soft collisions' have a $Q^2$ of a few GeV$^2$, hard collisions at the LHC
are typically (100 - 1000 GeV)$^2$.

\begin{figure}[hbt]
\includegraphics[scale=.080]{{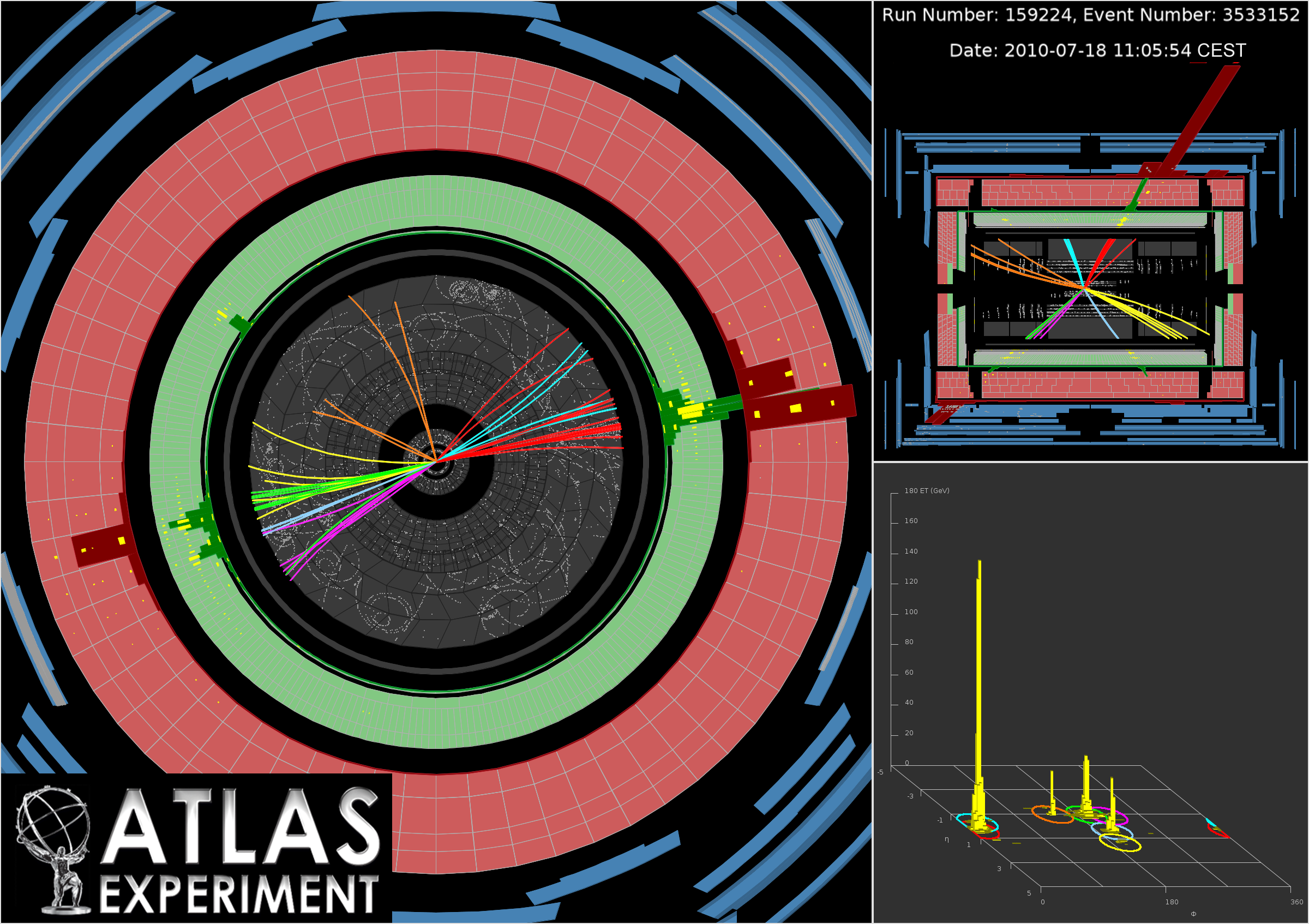}}
\caption{Two - jets event recorded at the LHC. The different components of the ATLAS detector are shown in the plane
perpendicular to the beam direction (left), along the beam (upper right) and in the plane of polar and azimuthal angles
(lower right).
The tracks in the inner detector and the energy deposition in the calorimeters are indicated and show that energy is
collimated in a rather small region, i.e. shows 'jetty' behaviour. 
\cite{bib:ATLAS-jets} } 
\label{fig:ATLAS-2jets}       
\end{figure}

The outgoing electrons of  Eq.~\ref{eq:hdecay} are rather easy to simulate.
More complicated are reactions where also the final state particles $C$, $D$ are strongly interacting partons.
Due to the specific properties of the strong interactions, these partons cannot be observed directly, but
are 'dressed' by many additional partons at distances smaller than 10$^{-15}$ m. 
In the detector hard partons appear as narrow cones of some 20 - 40 particles, deemed 'jets'. 
A jet event recorded with the ATLAS detector is shown in Fig.~\ref{fig:ATLAS-2jets} 
To understand this dressing is important to infer from the measurement on the initial parton collision.
It is modeled based on the precisely probed theory of strong interactions (Quantum 
Chromo Dynamics, QCD) and simulated starting from the high $Q^2$ parton collision  
Eq.~\ref{eq:ABCD}
to low $Q^2$, where QCD inspired models\footnote{Here the particle physicists' notion of 'model' is 
used, which refers to a theoretical description
of a physics process that is not fully calculable from the well founded and established 'theory' 
of the Standard Model.} of parton emissions are invoked. 
These models are not unambiguous and several variants exist, all being cast into computer codes
using Monte - Carlo methods. 
We will now provide more details on the individual steps.

\subsection{The Matrix Element of the Hard Collision}
\label{sec:ME-hard}

The fundamental differential equation for the process Eq.~\ref{eq:ABCD} 
is known, but cannot be solved in a closed form.
Instead it is perturbatively expanded in the strong coupling $\alpha_s(Q^2)$
leading to quantum mechanical matrix elements, which can be solved exactly. 
Schematically this can be written as (see eg. \cite{bib-Salam2011}) 

\begin{equation} \label{eq:pertexp}
\sigma (A+B\rightarrow S) \ = \ \sigma_0 (A+B\rightarrow S) \ + \ \alpha_s \sigma_1(A+B\rightarrow S) \ + \
                                                         \alpha^2_s \sigma_2(A+B\rightarrow S) \ + \ .....
\end{equation}

where $\alpha_s $ is given by Eq.~\ref{eq:as-MZ}. The higher the order in $\alpha_s$, 
the more complex is the calculation, but since $\alpha_s$ is of order 0.1 its contribution 
also becomes smaller. 
The complete perturbation series cannot be determined
and is instead truncated, for
LHC processes typically at the $\alpha_s^2$ term. 
I.e. small contributions to the full cross - section are missing.

\subsection{Parton Distribution Functions: Dressing the Initial State}

Since the partons $A,~B$ in the initial state are subcomponents of protons they
assume just a fraction of the proton energy $p$, parametrized as 

\begin{equation}
x(\mathrm{parton}) \ = \ \frac{p(\mathrm{parton})}{p(\mathrm{proton}) } \ \ \ \mathrm{for} \ \ p(\mathrm{proton}) \ \rightarrow \ \infty
\end{equation}

$x$ follows a probability distribution denoted as 'parton distribution function (pdf)', 
and enters the prediction of a cross section but cannot be calculated. 
The $Q^2$ dependence of the pdfs is, however, given by theory and has been experimentally
well confirmed (see Sect.~\ref{pdf-valid}).

\subsection{Dressing of the Outgoing Partons}
\label{sec:dressing}

To dress the outgoing partons $C$, $D$ into jets, QCD based models are invoked, in which
each final parton of the matrix element calculation is assumed to branch into further
partons, which subsequently branch again. 

The models follows each parton individually neglecting quantum mechanical interferences.
Each of these branchings happens at some $Q^2$, which decreases with the order of the
branching. Since $Q \ \sim \ 1/t$ ($t$ denoting time), this parton showering can be interpreted as a 
time ordered (Markov) chain. 
Modeling particle production in jets
can be classified into three steps (for more details see \cite{bib-Salam2011,bib-Seymour}). 

%
\begin{figure}[hbt]
\includegraphics[scale=.35]{{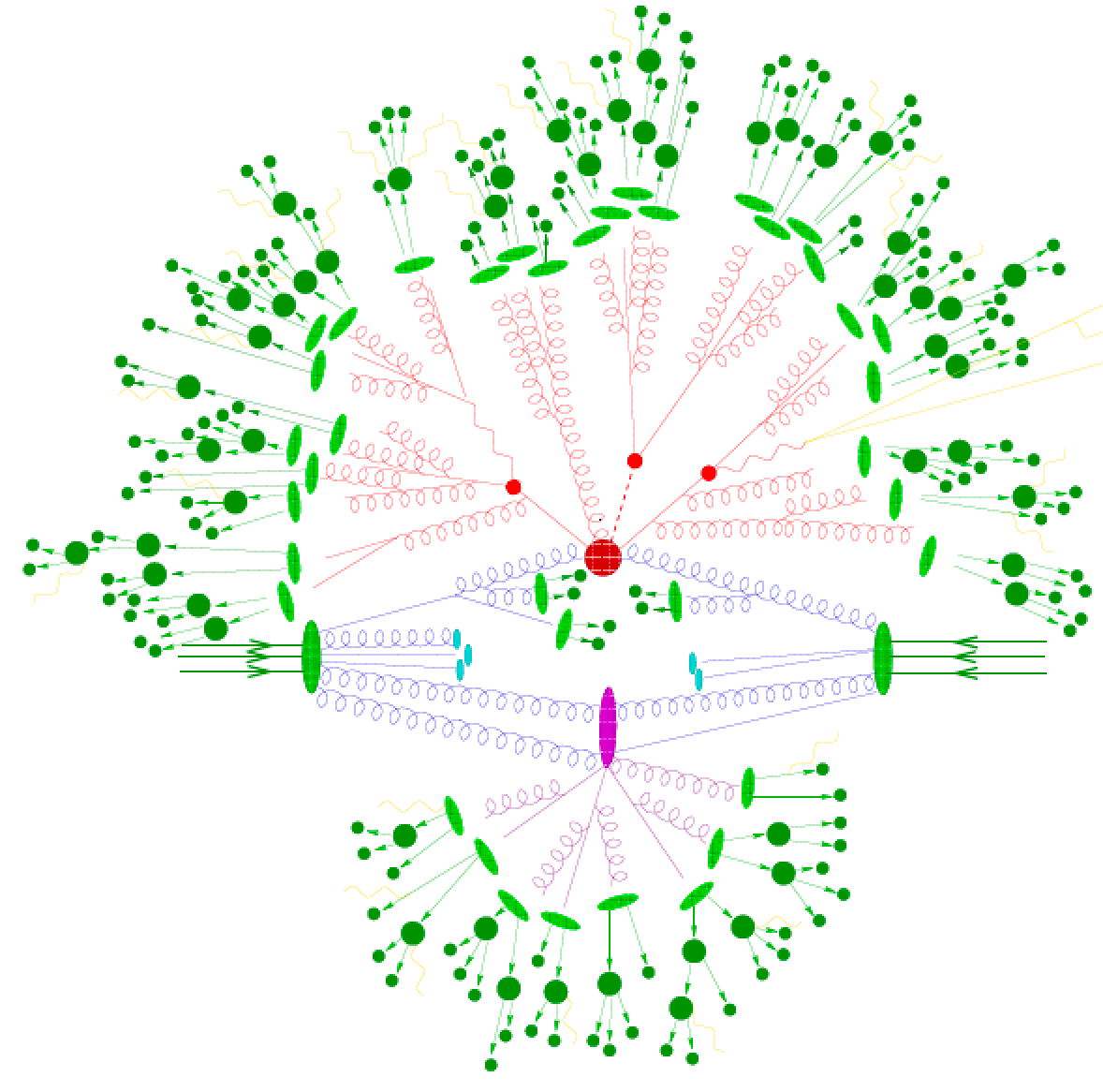}}
\caption{Schematic LHC event: straight lines denote quarks, curly lines gluons. 
The initial partons make a hard collisions (denoted by the large red circle) and split into
other partons. Four immediately outgoing
partons are shown, one (dashed line) decays only after flying a small distance. These partons branch
further with increasingly smaller energy and angle resulting in jet like structures.
Finally these partons turn into hadrons (shown as green
lines/dots). In parallel the other partons inside the protons also interact (elongated pink area) leading to
hadrons. 
\cite{bib:Schematic-tree} }
\label{fig:event-schematics}       
\end{figure}

\begin{enumerate}
 
\item The scattered partons $C,~D$ split into more partons, a so - called 'parton shower'. How the energy of a
         single parton is split among its daughters follows directly from theory. 
\item These partons are finally turned into stable hadrons making up 'jets'. The details of this 'hadronization' 
         are condensed in 'hadronization functions' expressing the probability of the energy and the kind of
         hadron to be produced. They are provided using previous measurements.         
\item Whereas in the hard process just one of the many partons inside each proton interacts, the remnant ones
         also produce a spray of particles. This 'underlying event' is measured using special LHC processes.

\end{enumerate}
                  
This means that jet evolution can be simulated with a Markov chain
by describing individual steps with special probability distributions. 
A rather complicated particle structure emerges, 
as schematically depicted in Fig.~\ref{fig:event-schematics}, where the many and diverse
parton branchings become apparent, motivating the use of Monte - Carlo simulation. 

Events at the LHC are further complicated by the occurence
of hadrons from additional $pp$ collisions in the same LHC bunches. These contribute
a 'pedestal' of particles to the hard collision of interest and are denoted as 'pile - up' events. 
These can be determined from data (see Sect.~\ref{sec:soft}).

\section{Detector Simulation}
\label{sec:DetSim}

\begin{figure}[hbt]
\includegraphics[scale=.30]{{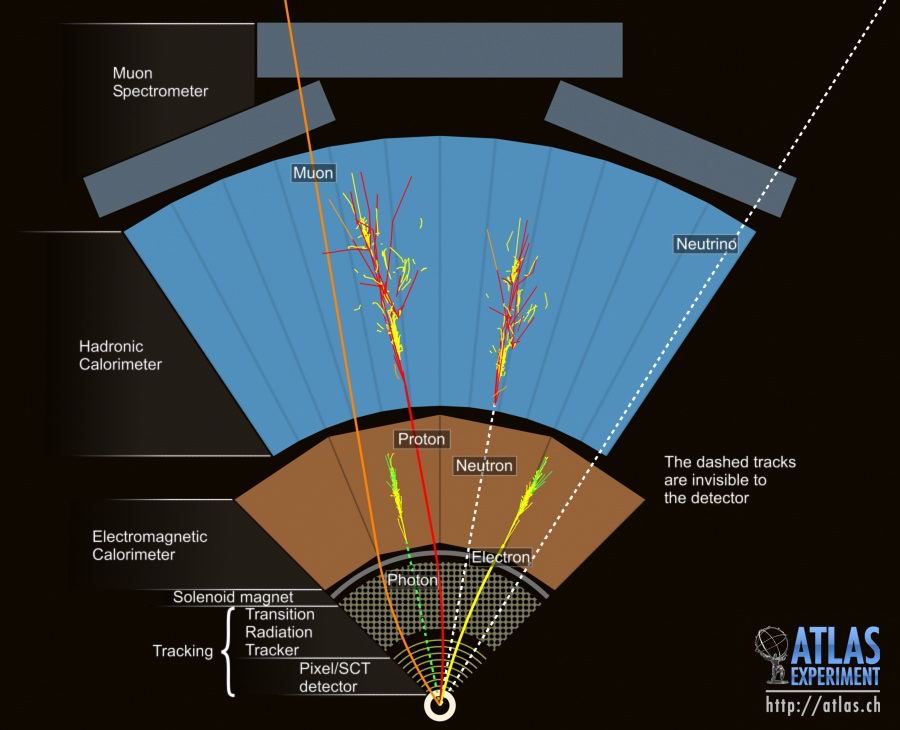}}
\caption{Schematics of the interactions of various types of stable particles in the components
              of a typical LHC detectors. As becomes apparent by combining the information the particle type
              can be identified \cite{bib-det-schematics} }
\label{fig:interactions}       
\end{figure}

The modeling of the physics process 
terminates with a list of stable particles:  hadrons, photons, electrons, muons and 
neutrinos. These are relevant for what is seen in the detector.
Apart from neutrinos all particles interact with the detector material and generate
electronic signals. Roughly speaking, as becomes apparent in Fig.~\ref{fig:interactions},
each of the components of the ATLAS detector has a special sensitivity to one of  these
particles. Their reconstruction provides a picture of what has happened at the
collision point.   

The principle of detector simulation
is that each stable particle is traced inside the detector 
up to a volume element containing some material. 
They may interact according to probabilities obtained from models.
The products of the interactions are then further traced to the next volume element with
material and so on\footnote{For a more detailed and also historical account of detector simulation in particle physics 
see~\cite{bib:Simulation-history}.}. 
All material affects the passage of a particle and has to be considered, but
only part is 'active' meaning that it generates electronic signals that are used to 
reconstruct the event.

These interactions and generated signals are cast into computer codes~\cite{bib:GEANT} 
that have been developed
over the last 30 - 40 years and applied and validated in various, rather different detector environments. 
The main ingredients of these simulations are

\begin{itemize}
\item the geometry and materials of the detector,
\item methods of numerical integration to follow a particle, partly inside a magnetic field,
\item modeling the particle interactions in the material.
\end{itemize}

\begin{figure}[hbt]
\includegraphics[scale=.13]{{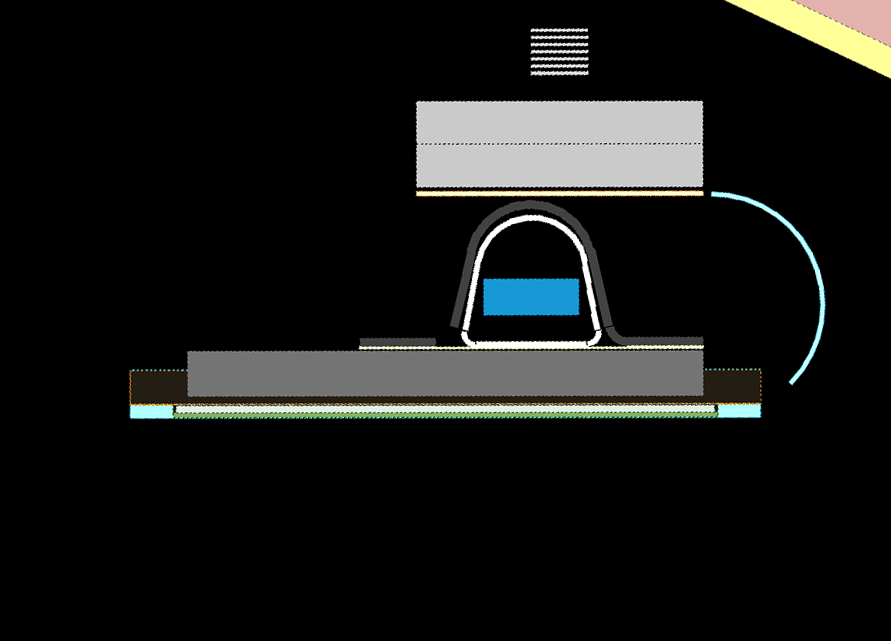}}
\includegraphics[scale=.165]{{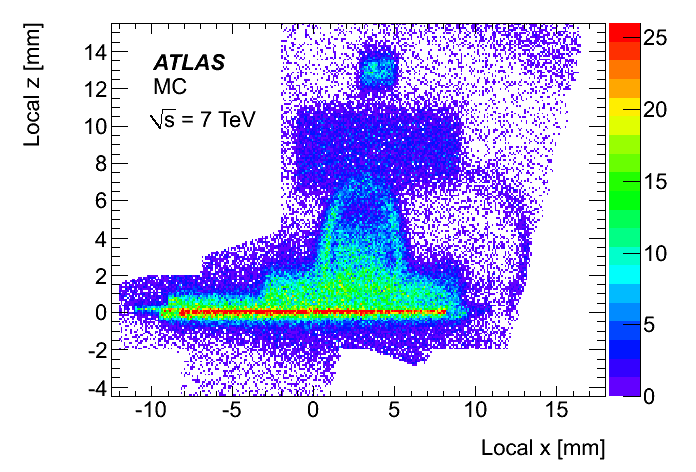}}
\includegraphics[scale=.165]{{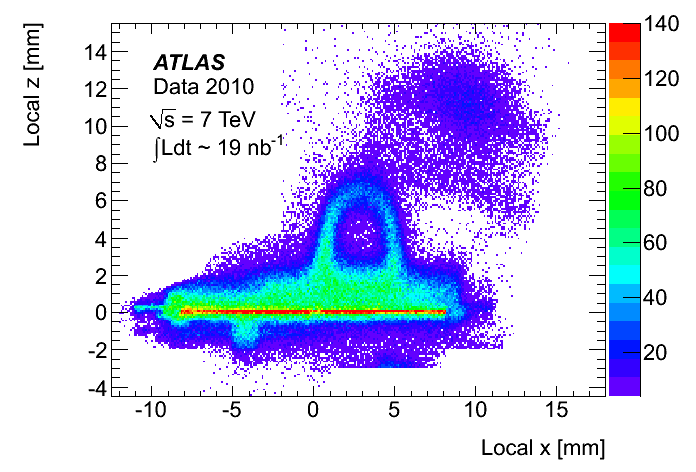}}
\caption{Display of a component of the ATLAS pixel detector. The total length of the object is 2cm, the widths of  individual
              parts are as small 0.02 mm. (a) The distribution of the material in the simulation 
              \cite{bib-det-schematics},  
              (b) the material in the simulation
              as seen in interactions of particles, (c) the material as seen in real collision events. Red indicates many interactions,
              blue relatively few \cite{bib-ATLAS-Tomography}.}
\label{fig:Pixeldetails}        
\end{figure}

The detector geometry is mapped in a first step according to engineering drawings that 
have been used for building
the detector. When relevant, one includes fine details, in part as small as 
several $\mu m^3$, as can be seen from Fig.~\ref{fig:Pixeldetails}(left) showing schematically
one of the 80,000 modules of the pixel detector, the innermost component of ATLAS. 
The interactions 
with the material are simulated stochastically using the interaction cross section of the 
incoming particle $h$ with the material $A$ under consideration. i.e. 

\begin{equation}
\sigma_{\mathrm{int} } (h+A )\ = \ \sigma_1(f_1) \ + \ \sigma_2(f_2) \ +  \ ......  
\end{equation}
 
where the $f_i$ are possible final states, which may consist of several particles. 
The respective
momenta are generated according to models cast into computer codes, e.g.~\cite{bib:EGS,bib:FLUKA}. 
Those for incoming electrons and photons are rather well understood, interactions
of hadrons are more uncertain.

The interactions in the active volume will then be digitized, i.e. translated into an electronic 
signal.

While the simulation should be as precise as possible, it is not meaningful to exceed the measurement 
uncertainty and also it should be balanced with the required computing time. 
Therefore some effects are integrated over and condensed into a single parameter. 
 
The simulation of the pixel detector is
an example of the balance between fine details and required precision.
There an electronic signal is caused by some 30,000 
electron - hole pairs that are produced from the passage of a particle through a 250 $\mu$m thick 
silicon layer.
It is well understood how these pairs are produced and move towards the electrodes, thus 
simulating these would be possible. However, such detailed simulation is only 
meaningful if supplemented with a simulation of each of the somewhat different
80,000 electronic circuits, requiring an excessive use of computing resources.
Instead the response of the pixel detector to particles is measured and the probability distribution of
the signal depending on the particle's properties is used in the simulation without considering
the details of its generation. No uncertainty is induced by this discretization. 

As a side remark, one should be aware of the hugely different scales in detector simulation. Within a
detector of a global diameter of 25 m, structures of the size of 0.00001 m are considered, if important. 
How finely the volume elements in the simulation are granulated, depends one their impact on the 
measurement and has been tested. 

The simulated electronic signals in the various parts of the detector
are then subjected to the identical procedure as the recorded data
to reconstruct physics objects, i.e. electrons, muons, jets etc..

\section{Principles of Validation and Uncertainties}
\label{sec:valid}


Formally simulation in particle physics
translates an original, 'true' distribution $T $ of partons $C,~D$ with energy $E_{C,D}$, 
3 - momenta $\mathbf{p}_{C,D}$ and types $f_{C.D}$ into $n$ reconstructed particles with
respective energies, 3 - momenta and types, i.e.

\begin{equation}
\left(
\begin{array}{c}
E_1  \\ \mathbf{p}_1 \\ f_1\\ E_2 \\ \mathbf{p}_2 \\ f_2 \\.... \\ .... \\ E_n \\ \mathbf{p}_n \\ f_n
\end{array} 
\right)
  \ = \ 
{\cal{M}} \times   
\left(
\begin{array}{c}
E_C  \\ \mathbf{p}_C \\ f_C \\ E_D \\ \mathbf{p}_D \\ f_D
\end{array} 
\right)
\end{equation}

where $n$ is ${\cal{O}}$(1000 - 100000). The transformation  $ {\cal{M}} $ expresses what is happening 
in the simulation.
Instead of listing all momenta of individual particles, the main idea can be visualized in
a simpler way by considering some distribution $z$ of (physics) interest, e.g. the mass $M$ of 
an event. 

Along the discussion of Sect.~\ref{sec:ME-hard}, the matrix element calculation in some model
yields the 'true' mass distribution $T(M_{C,D})$. 
In simulation the $C,~D$ are dressed and affected by experimental resolutions
such that $T(M_{C,D})$ is transformed into a prediction $P(M)$ for
the mass distribution that is supposed to be measured for the model under consideration. 
Actually measured in the experiment would be a distribution $D(M)$.
To interpret the measurement, e.g. to infer on the validity of the model, $P(M)$ 
has to be compared to $D(M)$.

\subsection{Factorization of Migration}
\label{sect:factorization}

In terms of $z$ simulation means

\begin{equation}
P(z) \ = \ {\cal{M} } T(z)
\end{equation}

Discretizing $z$ in intervals
the transformation
can be expressed by a square matrix of the migration of a theoretical mass value to an observed one.

\begin{equation}
{\cal{M} } \ = \ \mathbf{m_{ij} } \ = \ 
\left(
\begin{array}{c c c c}
m_{11} & m_{12} & . ....& m_{1n} \\
m_{21} & m_{22} & . ....& m_{2n} \\
 .....&  .....              &  .... &  .....  \\
m_{n1} & m_{n2} & . .... & m_{nn}
\end{array}
\right)
\end{equation}
 
where each element in the matrix relates a value of $M$ in the incoming step to a mass of the outgoing
steps. E.g $m_{2n}$ is the probability that an event with a true $z$ in bin $n$ has a reconstructed
$z$ in bin $2$.
In the previous section it was discussed that simulation proceeds in (time - ordered) steps. Using the
formalism each step corresponds to a
special migration matrix ${\cal{M}}_{\alpha }$.
Here $\alpha $ denotes the effect under consideration, e.g. $\alpha $ = 1 the effect of the pdfs, 
$\alpha $ = 3 the one due to hadronization,   
$\alpha $ = k  is the distortion from a detector component etc.. 
The complete simulation ${\cal{M} }$ is then factorized into ${\cal{M}}_{\alpha }$,

\begin{eqnarray}\label{eq:factorization}
{\cal{M} } \ & = \ \left[ {\cal{M}}_{k } \times .....\times {\cal{M}}_{2} \times {\cal{M}}_{1} \right]  \\
P(z) \ & = \ \left[ {\cal{M}}_{k} \times ...... \times {\cal{M}}_{2} \times {\cal{M}}_{1} \right] \times T(z) 
\end{eqnarray}

More precisely, adding pile up $U(z)$ and background processes $B(z)$ to the signal process $S(z)$,
one arrives at modified (primed) results.

\begin{eqnarray}\label{eq:TSBU}
T(z) \ \rightarrow \ T'(z) \ & = \ (S+U)(z)  \  + (B+U)(z) \\ 
P'(z)  \ & = \ {\cal{M} } \cdot \left[  (S+U)(z)  \  + \ (B+U)(z)  \right]
\end{eqnarray}

The physics question is whether $P'(z) \ = \ D(z)$, which means $S$ 
is correct, if ${\cal{M} }$, $U(z)$ and $B(z)$ are correct, but can also be fulfilled if e.g. $S$ and ${\cal{M} }$ 
are incorrect. 
Inferring on $S$ from $D$ requires one to validate ${\cal{M} }$ and the background distributions.

\subsection{Is Factorization correct?}

These considerations, which will be crucial for the validation procedure
discussed in the next sections, 
depend on factorization.
It is evidently fulfilled if the steps are incoherent and follow a time sequence. 
This is clearly true for the detector simulation, where a particle from
the $pp$ interaction first interacts in the tracking detector before entering the calorimeter etc..
I.e. only those particles have to be considered in the simulation of the calorimeter response
that leave the tracking detector (cp. Fig.~\ref{fig:interactions}).  

The time - sequence in the transition from the matrix element to the final
hadrons in the physics generator, however, is only approximate since quantum level interferences
are left out. However, the factorization assumption is justified from theoretical arguments 
(e.g.~\cite{bib-Salam2011})  at least for pdfs to rather high precision. 
The basic argument is that
the different steps occur at significantly different energy scales, which hardly influence each other.
Similar arguments for the validity of factorization can be applied to the showering and hadronization effects. 
In addition the assumption has been tested by 
comparing processes involving electroweak particles\footnote{I.e comparing a $W$
decay in $e\nu$ with those in $q\bar{q}$ or production cross sections and jet structures in
of $e^+e^-\rightarrow q\bar{q}$}. 

\section{General Procedures of Validation in Particle Physics}
\label{sec:genproc}

At face value data and simulation agree for SM processes at the LHC to stunning precision.
In so far, the requirement in the philosophical literature, 'validation .... is said to be the 
process of determining whether the chosen model is a good enough representation of the realworld
system for the purpose of the simulation' \cite{bib:EnzSimu} seems to be fulfilled.
However, as outlined above, the agreement is a necessary,
but not a sufficient condition to claim the correctness of the model of the underlying parton process. 
Instead the migration matrix ${\cal{M} }$ and the backgrounds have to be validated. The principle 
validation methods will be discussed in the next sections. The basic ideas are as follows.

\begin{enumerate}
\item For each ${\cal{M}}_{\alpha }$ the corresponding model is validated by either data from previous experiments,
         possibly together with a well founded theory, or better even, by an in - situ experimental validation. 
         For this, dedicated processes are used, each isolating a specific effect.
         If all steps of the simulation are validated then, obviously,
         the whole simulation is validated and the sufficient condition is fulfilled to infer from the agreement
         of simulation and data on the underlying physics process. All steps means that both the underlying
         physics model, which is a genuine part of the simulation, and the description of the detector response
         have to be validated 
\item  Given that simulation should help to provide quantitative statements on SM parameters or 
          BSM effects, 
          validation should provide an estimate of how well a process is validated. 
          This is expressed as a 'systematic' uncertainty for each validation step $\delta ({\cal{M}}_{\alpha } )$. 
          In this sense the process of validation is synonymous with the process of assigning
          an uncertainty.
          By error propagation this leads to a total uncertainty $\delta (P'(x))$ of the simulation prediction.         
\end{enumerate}


It is beyond the scope of this article to cover the complete validation procedure, which is
documented in numerous publications and notes (some will be mentioned in the text).
Instead some examples will be presented in more detail, highlighting general methods 
used in the validation of both physics and detector modeling

\begin{itemize}
\item a combination of previous experiments and model assumptions,
\item measurements of LHC processes that are complementary to the process of interest, 
\item in-situ measurements of properties of the detector,  
\item adjusting the simulation to data using previous precision measurements.       
\end{itemize}


As will hopefully become apparent in the following sections, the different 'factors' that enter simulation
at the LHC are experimentally tested and simulation is applied on a significantly constrained 
material basis.  This discussion,
although maybe sometimes technical, seems to be important in view of claims that 'in the context
of the LHC .... there is a lack of experimental data for comparison [with simulated data]'
\cite{bib:Morrison}(290).

\section{Validation of the Physics Generators}

The parton distributions in a space - time region of smaller 10$^{-15}$ m and the model
of how they turn into hadrons, have been discussed in Sect.\ref{sec:phys-mod}.
These processes are statistically distributed and thus a generic part of the computer simulation. 
Moreover, since the parton evolution is described by models, these have to be validated.

\begin{figure}[hbt]
\includegraphics[scale=.65]{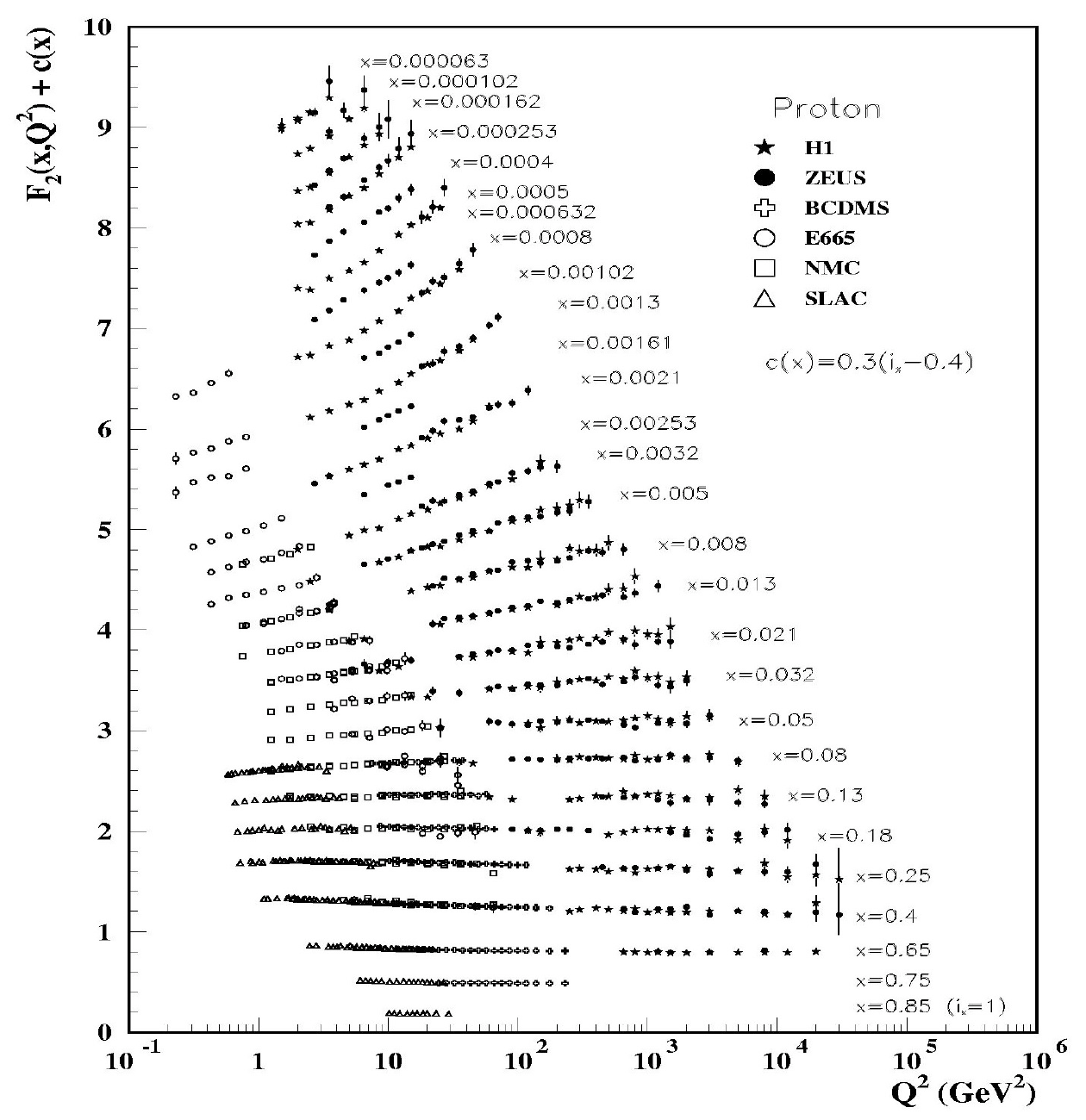}
\caption{Measurements on the $Q^2$ dependence of the pdf. Shown are measurements
from several experiments and in different bins of $x$ \cite{bib-PDG}. }
\label{fig:Q2-dependence}       
\end{figure}

As examples validation of pdfs and pile - up events will be discussed in the following sections. 
But before a brief comment on the other steps. As discussed, the perturbation series of the matrix element of
the hard process are truncated. The magnitude of the remaining terms is estimated by a convention
that has been tested in various processes such that uncertainties of typically some 3-5\% are assigned.
The basic understanding of the major other steps, i.e.
showering and hadronization (see Sect.~\ref{sec:dressing}), has been obtained from previous 
experiments at $e^+e^-$ colliders  \cite{bib:Maettig, bib:LEPfrag} and cross - checked 
with LHC data (see e.g. \cite{bib-LHC-showering}). The steps  are described by various
models, and their differing outcomes provide the uncertainty range assigned to showering effects.

\subsection{pdfs}
\label{pdf-valid}

The pdfs are an essential ingredient of the simulation and instrumental for the interpretation of 
measurements. 
The cleanest way to measure pdfs is via the scattering of electrons on nucleons.
A compilation of the major experiments is shown in Fig.\ref{fig:Q2-dependence} 
for various intervals in $x$ as a function of $Q^2$.
Whereas the pdfs themselves cannot be calculated from
first principles, their $Q^2$ dependence is precisely known in QCD and excellently confirmed by
measurements.
Knowing the pdf at one $Q^2$ allows one to extrapolate it to another $Q^2$ 
(for more details see~\cite{bib-pdfreview}).

%


Such extrapolations are needed for the LHC, for which much higher $Q^2$ can be reached.
Validation of the LHC has to address the question: what variations of pdf models are allowed 
by data at low $Q^2$? Given the knowledge of the $Q^2$ dependence how large are the
variations at LHC conditions? Are there ways to test pdfs directly at the LHC avoiding
circularity?

The uncertainties of pdfs at a certain $Q^2$ are due to both
measurement uncertainties and theoretical ones. The latter ones come 
e.g. from inter- and extrapolating the binned measurements to a continuous distribution. 
Given the number of input bins there is in general only a small amount of variation allowed. 
However at the extremes like $x$ = 0, 1, they can be important. 
These ambiguities lead to different pdf models with some variation in the expectations for the LHC,
none of which can be a priori excluded. 

\begin{figure}[hbt]
\includegraphics[scale=.50]{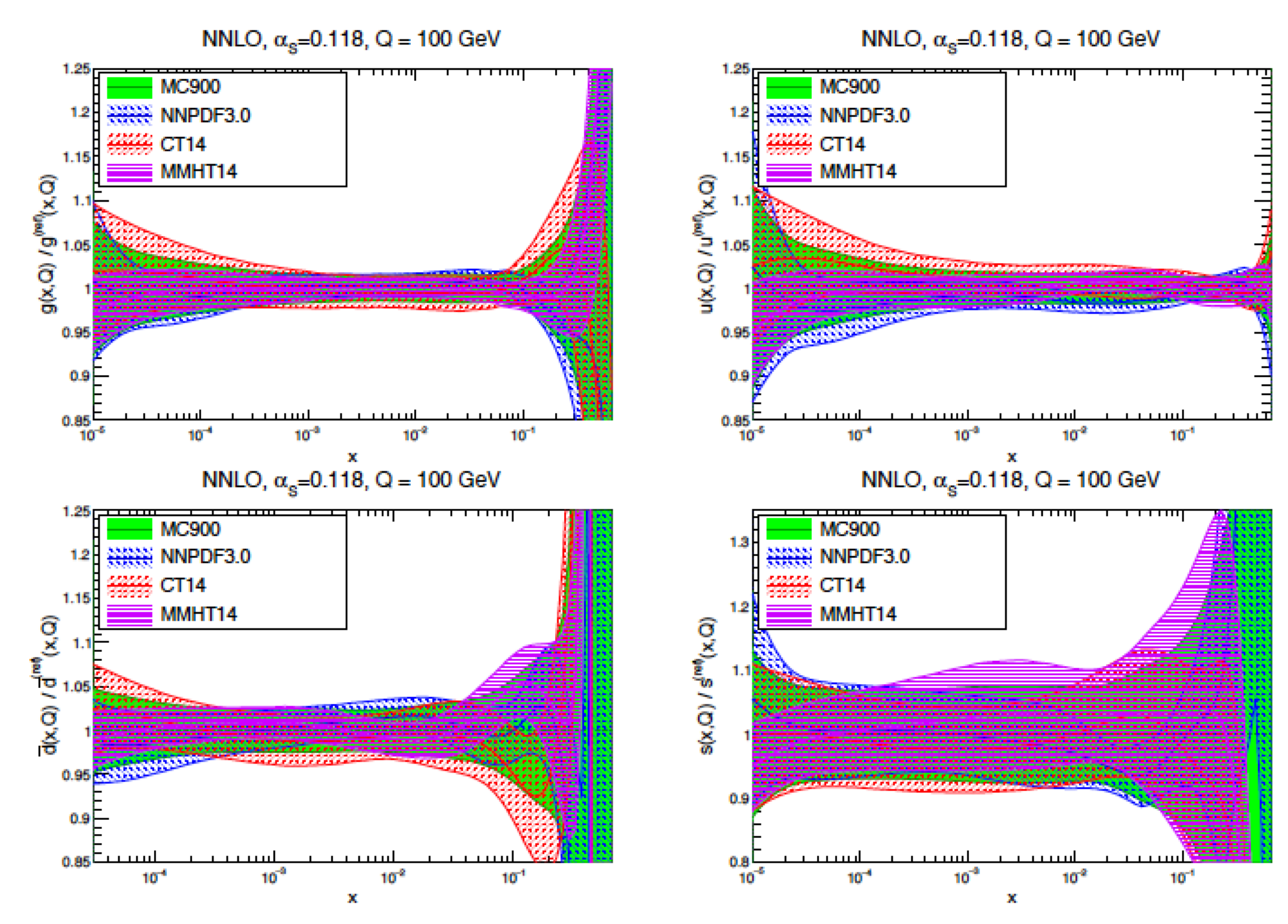}
\caption{Predictions for the pdfs of gluons (upper left), up quarks (upper right),
anti - down quarks (lower left) and strange quarks (lower right) are shown for $Q$ = 100 GeV
and an $x$ range stretching over 5 orders of magnitude. The different colours correspond to different
pdf models, all are shown being normalized to the average of these. The coloured bands correspond to
the uncertainties assigned to the corresponding model. 
\cite{bib-pdf-uncer} }
\label{fig:pdf-unc}       
\end{figure}

As an example, four different pdf models for different kinds of parton species 
at a $Q$ = 100 GeV (i.e. close to the relevant scale for the Higgs production)
are shown in Fig.~\ref{fig:pdf-unc}.   
As can be observed, the difference between different models is in general smaller than the uncertainty
assumed for an individual model - an indication that the constraints from data are fairly tight. 
The variation between these pdf models together with their
individual uncertainties are used as an overall pdf uncertainty of typically below 10\%. A notable
exception are very high $x$ values.
    
\begin{figure}[hbt]
\includegraphics[scale=.085]{{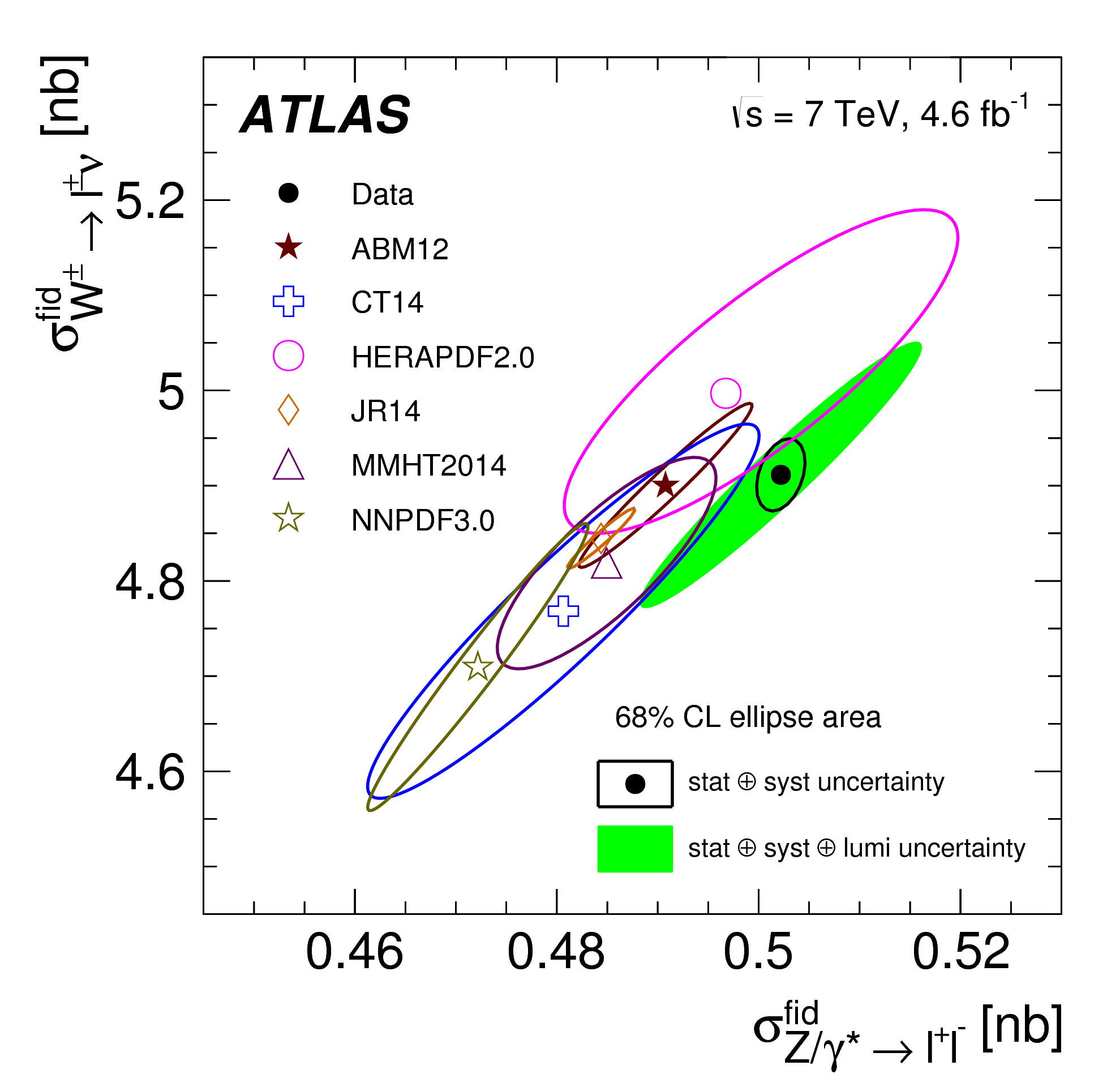}}
\includegraphics[scale=.085]{{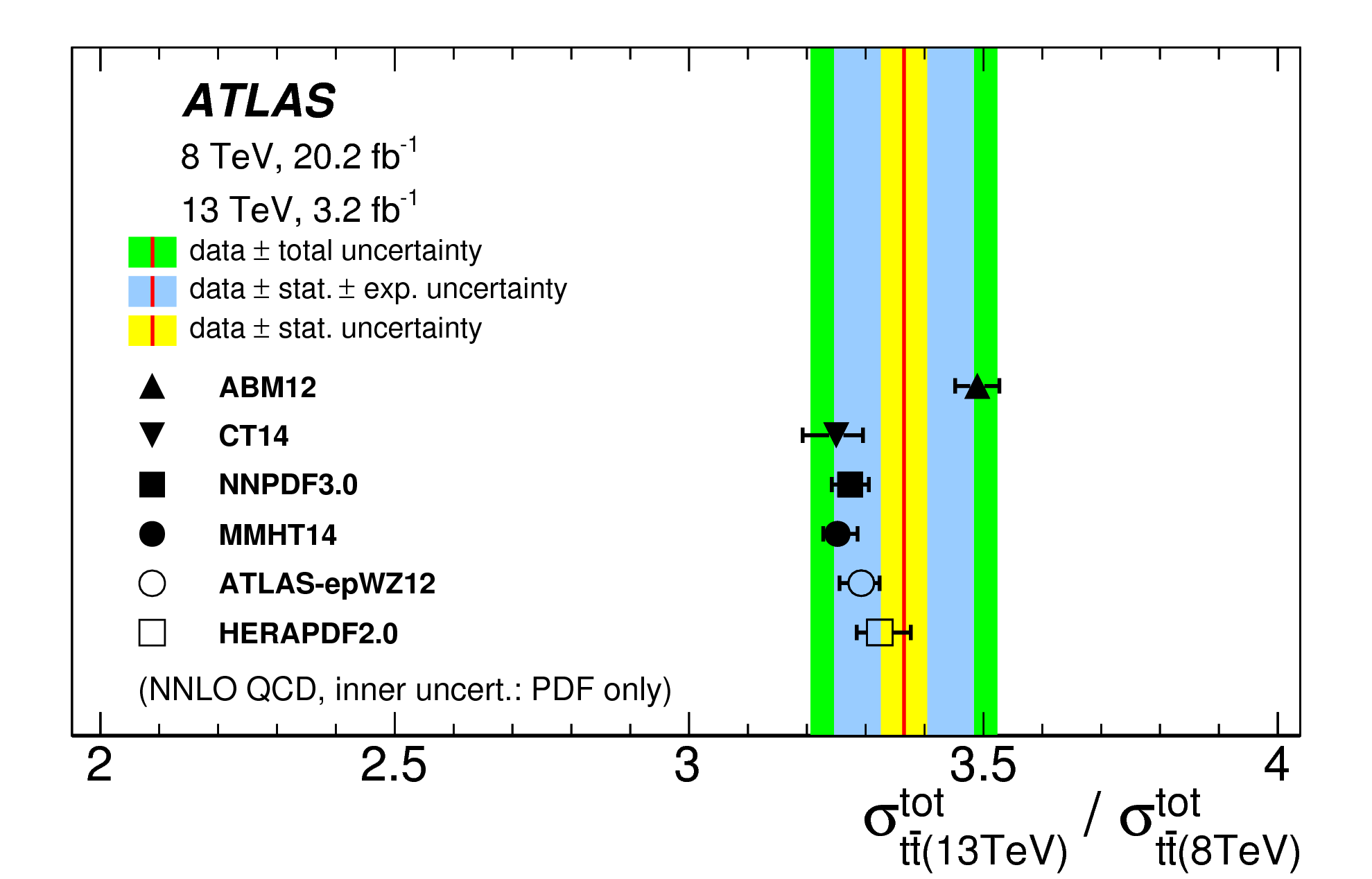}}
\caption{ATLAS Measurement of 
              $W^{\pm}$ vs. $Z^0$ (left), being sensitive
              sensitive to other quark species. The data are compared to the expectation from various
              pdf models. The ellipses indicate the corresponding uncertainties, i.e. the 66\% certainty range.
              \cite{bib:ATLAS-pdf}. 
              The right figure compares production of pairs of top quarks at two different
              LHC energies \cite{bib:ATLAS-tt-pdf}.}
\label{fig:ATLAS-pdf}        
\end{figure}

In addition, these predictions can be validated by using special LHC processes.
Particularly those involving $W$ and $Z^0$ production provide in situ constraints on pdfs
\cite{bib:ATLAS-pdf}. 
In addition the energy dependence of the yields of theoretically well understood processes 
like the production of top quarks ($t\bar{t}$) can be used~\cite{bib:ATLAS-tt-pdf}. 
Both examples are depicted in Fig.~\ref{fig:ATLAS-pdf}.
These cross checks show that the assumed pdf models derived from non - LHC 
experiments and their uncertainties are consistent with the LHC measurements.   

In addition, these predictions can be validated by using special LHC processes.
Particularly those involving $W$ and $Z^0$ production provide in situ constraints on pdfs
\cite{bib:ATLAS-pdf}. 
In addition the energy dependence of the yields of theoretically well understood processes 
like the production of top quarks ($t\bar{t}$) can be used~\cite{bib:ATLAS-tt-pdf}. 
Both examples are depicted in Fig.~\ref{fig:ATLAS-pdf}.
These cross checks show that the assumed pdf models derived from non - LHC 
experiments and their uncertainties are consistent with the LHC measurements.

\subsection{Pile Up in pp Scattering}
\label{sec:soft}

As stated in~\ref{sec:dressing}, every hard interaction of interest is accompanied by some 30 additional and incoherent
$pp$ interactions in one bunch crossing, called pile - up. This 'pile - up' has to be 
an integral part of the simulation of a LHC event. 
Validation of these requires the properties of these events to be understood.
This is achieved by direct LHC measurements (e.g. \cite{bib-minbias}).

Since pile - up events have a high cross section at the LHC, they are
frequently produced in bunch crossings, but can also be measured individually with 
short special LHC runs. Based on these measurements, QCD inspired 
models are devised to describe pile - up events. These models are then used 
in the simulation to overlay pile - up to the physics processes of interest.  
 
\begin{figure}[hbt]
\includegraphics[scale=.175]{{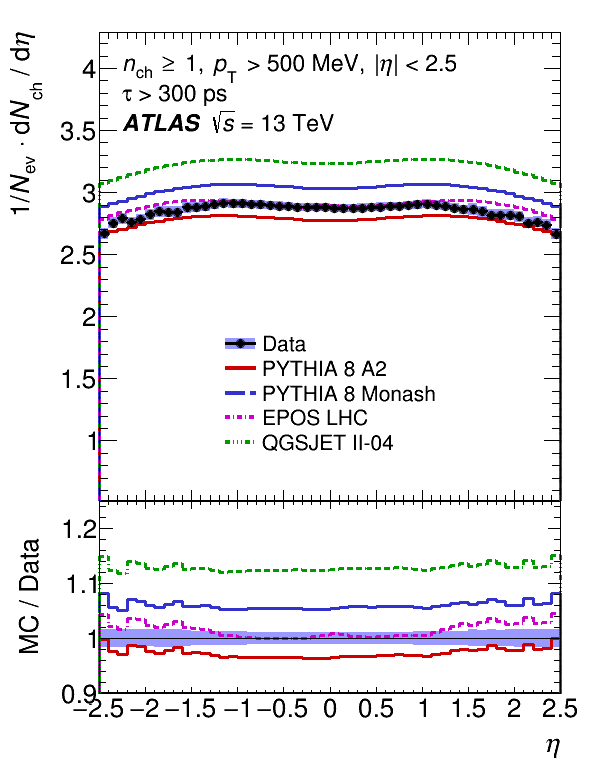}}
\includegraphics[scale=.055]{{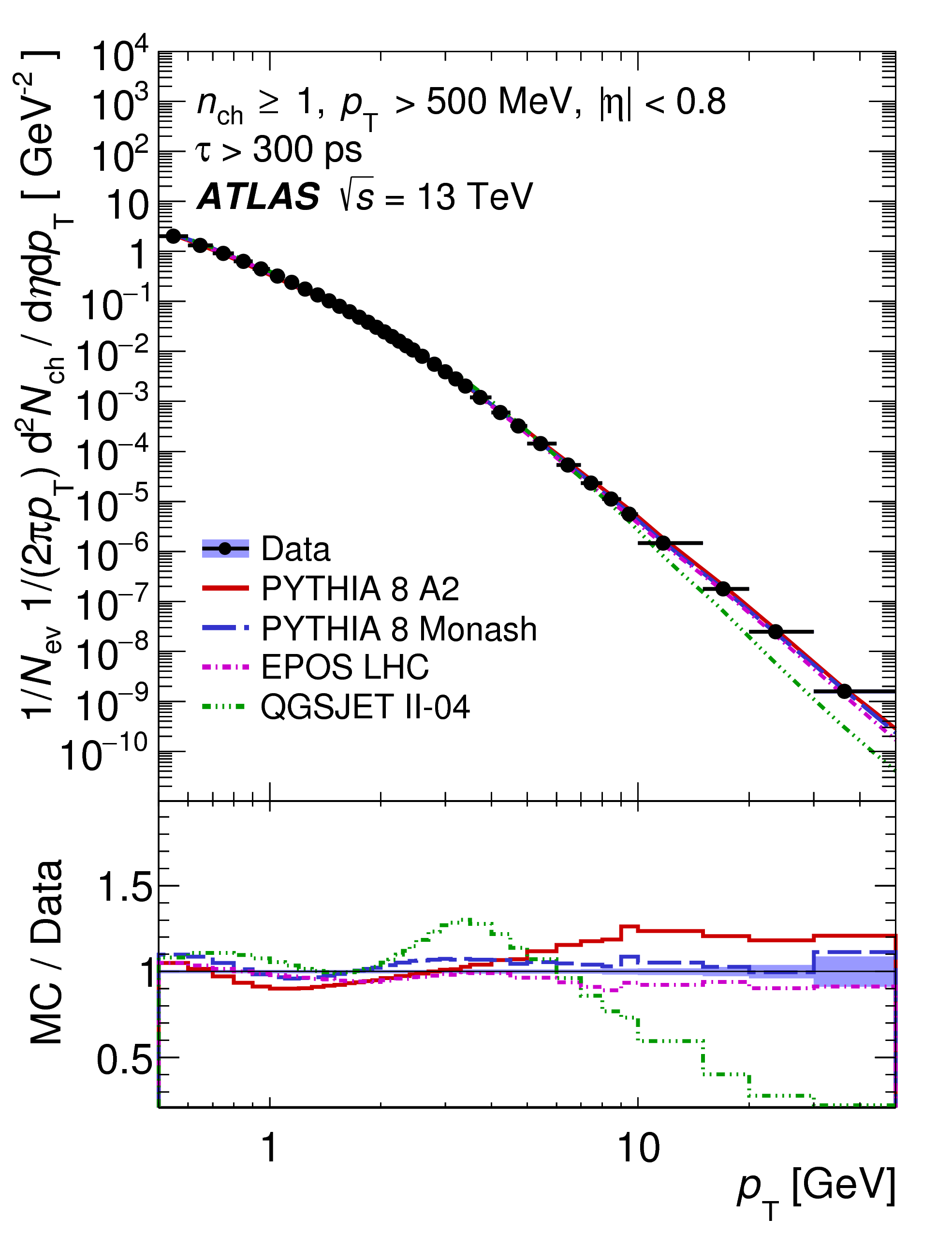}}
\includegraphics[scale=.175]{{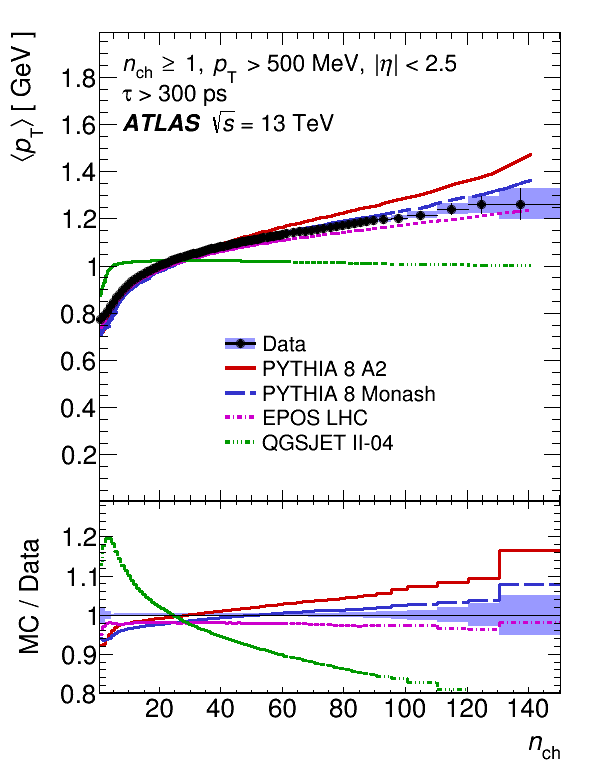}}
\caption{Relevant properties of one pile - up event as measured in low luminosity runs. The production angle wrt. the beam
              direction (left) and the transverse momentum $p_T$ distribution of charged particles are shown (center) and the average 
              event $p_T$ as a function of the number of charged particles in an event (right)
              \cite{bib-minbias} }
\label{fig:Minbias}        
\end{figure}

Pile - up events are found to largely produce an isotropic spray of low energy particles
as is apparent from the measurements shown in Fig.~\ref{fig:Minbias}..
In the Figure data are compared to several models. 
As can be seen some of them describe the measurements rather well and are 
used in simulation.
The remaining deviations between the preferred models and data are
small and do not affect the measurement in any relevant way.

Pile up events can be seen as an example, where the large variety of LHC processes can
be used to measure some process $i$ directly as input to simulation for complementary 
physics processes $j$.

\section{Validation of Detector Simulation}
\label{sec:val-det}

The modeling of the detector is crucial to understand how well particles can be 
reconstructed from electronic signals in the detector components  (cp.~Sect.~\ref{sec:DetSim}).
Here we focus on detector geometry and simulation of electron properties.

\subsection{Testing the Detector Geometry}

While the starting point for the description of a detector are engineering drawings, once 
the detector is installed details may change rendering the simulation inaccurate.
As an example, consider a module of the pixel detector whose engineering 
drawing is shown in Fig.~\ref{fig:Pixeldetails}(left) and serves as a blueprint for
the simulation.
Data allow one to 'see' the material distribution with a tomography - like method.
  
The method is based on the number and position of particle interactions inside the 
detector\footnote{These interactions in the detector 
are completely distinct from those $pp$ interactions to test the SM and find BSM signals.}.
Their frequency is a measure of the material.
For a pixel module the number of interactions in a projected area of 100$\times $100 $\mu$meter$^3$ 
is shown in Fig.~\ref{fig:Pixeldetails}(centre)\cite{bib-ATLAS-Tomography}.   
The one-to-one correspondence with the input geometry clearly shows the 
method to work\footnote{This test of validation tools represents another use of simulation in particle physics, 
mentioned in Sect.~\ref{sec:roleSim}}.

The interesting step is to apply this tomography to data. 
The result is shown in Fig.~\ref{fig:Pixeldetails}(right).
Whereas the general structure agrees well with the simulation, there are
also important differences. E.g. the rectangular component in the simulation around z $\sim$ 10 cm
correspond to cables, which are much more spread out in the real detector, the circular shape around
$x$ = 3 cm is a cooling pipe. In the simulation it is considered to be mostly filled with a liquid, in the
data it becomes apparent that most of it is in gaseous state. Furthermore at $x$ = -4 cm an electronic
component (capacitor) is visible in the data, which had been omitted in the simulation. 
After these differences became known, the simulation has been adjusted.

The details of this discussion are not important, but the message is that the geometry of the detector can
largely be checked and corrected with the data themselves.

%
%

A variety of procedures are used for different components.
They are examples of in - situ measurements 
based on well established types of reactions and redundancies in the detector.

\subsection{Validation of Electron Simulation}
\label{sec:E-depo}

A key particle for LHC physics is the electron, which is produced in many SM and perceived
BSM processes and relatively easy to identify. Data analysis requires one to know the 
detection efficiency (cp. Eq.~\ref{eq:eff}), the 'true' energy $E_{\mathrm{true}}$ and
 the energy resolution $\sigma_\mathrm{E}$ as given by 

\begin{eqnarray}
E_{\mathrm{meas}} \ = \ \left( 1 + \alpha \right) E_{\mathrm{true}} \\
E_{\mathrm{true}} \ \rightarrow \ f(E) \ \propto e^{(E-E_{\mathrm{true}})^2/2\sigma_\mathrm{E}^2 }
\end{eqnarray}

where $1 + \alpha $ is the electron energy scale indicating a possible mismeasurement,
and $\sigma_E $ is a measure of how strongly the measurements fluctuate.
Evidently it is crucial for the simulation to describe these appropriately.

\begin{figure}[hbt]
\includegraphics[scale=.08]{{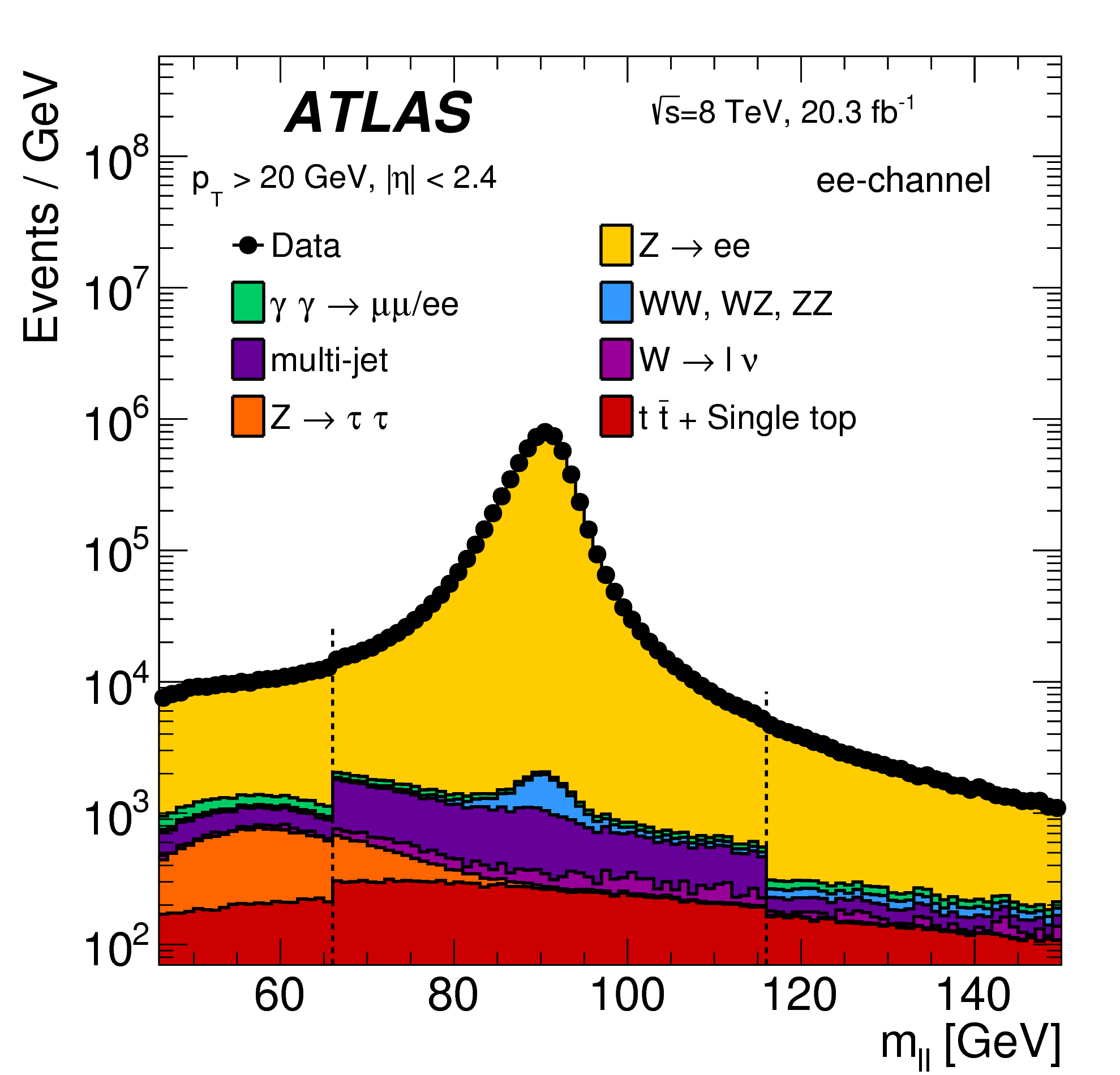}}
\includegraphics[scale=.09]{{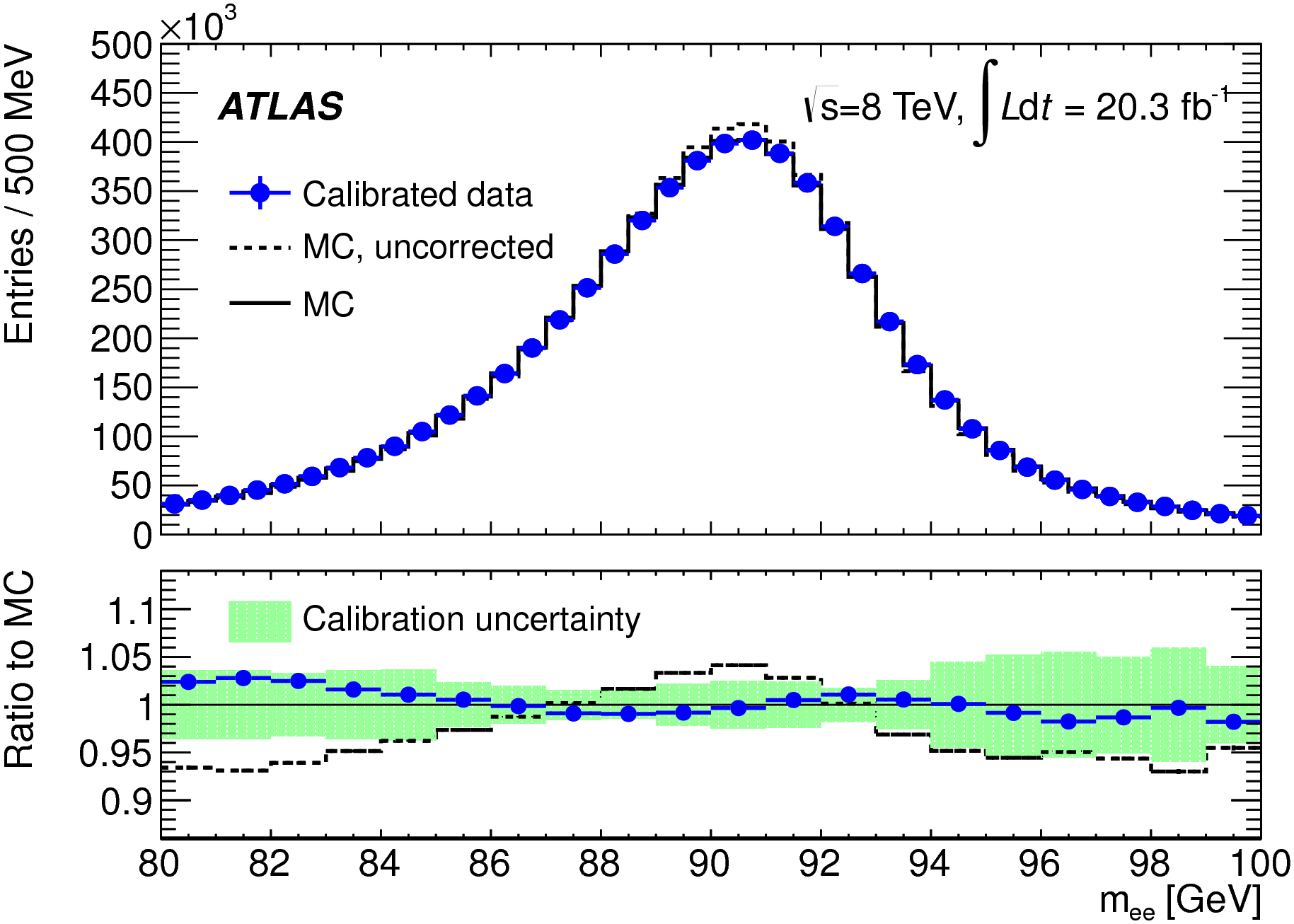}}
\caption{The $e^+e^-$ mass distribution observed with the ATLAS detector (left) \cite{bib:Z-measurement}. 
Also shown are the background contributions.
              Comparison of data and simulation before and after adjusting the simulation to the $Z^0$ peak as a function of the $e^+e^-$ mass (right)~\cite{bib:e-calibration}. The ratio between data and simulation is shown in
the lower part.
}
\label{fig:Z0}        
\end{figure}

Simulations of interactions of electrons with material are based on a well established theory, and  
have been studied numerous times and with many
detectors, also in test - beams for the calorimeter components later installed in ATLAS. 
The electronic response in the calorimeter to the electron energy, however, needs to
be calibrated. 
Given the precise knowledge of the $Z^0$ mass from previous
measurements at the $e^+e^-$ collider LEP \cite{bib:LEPZ0}, see
Eq.~\ref{eq:as-MZ}, and the abundant and clean $Z^0$ production at the LHC 
(see Fig.~\ref{fig:Z0} (left) \cite{bib:Z-measurement}), the electron energy is
calibrated to reproduce the $Z^0$ mass. 
To take into account distortions of the shape of the $Z^0$ peak due to secondary 
interactions of the electron with the material in front of the calorimeter, this calibration 
is based on the $Z^0$ shape obtained from simulation. 

Before this general procedure is realized, the simulation
of the electron inside the calorimeter has to be validated in - situ. E.g.
inhomogeneous material distributions are obtained from the longitudinal evolution of
a shower, and modifications of the calorimeter geometry due to gravitational effects
are derived from angular modulations of the ratio of redundant measurements in the
tracking detector and the calorimeter. 
The observed deviations are taken into account by parametric corrections of the 
simulation~\cite{bib:e-calibration}.

Whereas the $Z^0$ mass has been measured and can serve as a reference, the 
energy resolution is detector specific. Still, the $Z^0$ serves as a marker for validation in that its 
observed width is a measure of $\sigma_E$.
Differences between data and simulation are actually observed and the simulation
is smeared to accomodate this discrepancy. The impact of this correction in the 
simulation can be seen in Fig.~\ref{fig:Z0}(right), where the ratio of data and simulation of $m_{ee}$
is shown before and after the correction.  
Similarly the electron efficiency is determined by in-situ measurements of the $Z^0$ decay.
The simulation is adjusted to agree with the data~\cite{bib:e-eff}. 

I.e the simulation of electrons can be validated in - situ, making use of the redundancy of the
detector, the wide range of LHC processes and in particular a highly precisely measured 
reference process. The high statistics and precise knowledge of the references also imply that
the uncertainties assigned to simulation is small. Similar procedures also exist for other 
particles, e.g.~\cite{bib:mu-scale, bib:JES-Etmiss}.

\section{How Simulation is applied in Data Analysis} 

In this section two examples will be discussed of how simulation is applied in 
data analyses. These will also show methods to validate the background simulation.
In the first example properties of a known
particle, the Higgs boson, are determined in a kinematic region that is well probed. 
The second example is a search for
a BSM effect which requires extrapolations into as yet unexplored energy regions. 

\subsection{Measurement of the Higgs Cross Section}
\label{sec:Higgs}

In July 2012 both the ATLAS
and the CMS collaborations announced the observation of a new particle with a mass of
125 GeV \cite{bib:Higgs-discovery} through its decay into pairs of particles, especially
into $ZZ^*$ and $\gamma \gamma $.
To establish if this is the long sought for Higgs boson, the properties of the particle had to be determined. 
One key measurement is the cross section (cp. Eq.~\ref{eq:Xsec})  $\sigma_{h\rightarrow XX } $
for the Higgs production and its decay into two particles $XX$. 
Since the SM and alternative models for
mass generation lead to different cross sections, their precise measurements can discriminate them.
Here we will discuss $X\ = \ Z,Z^*$ along~\cite{bib-ATLAS-hZZ}, with each $Z$ decaying into
a pair of electrons. 

\begin{figure}[hbt]
\includegraphics[scale=.10]{{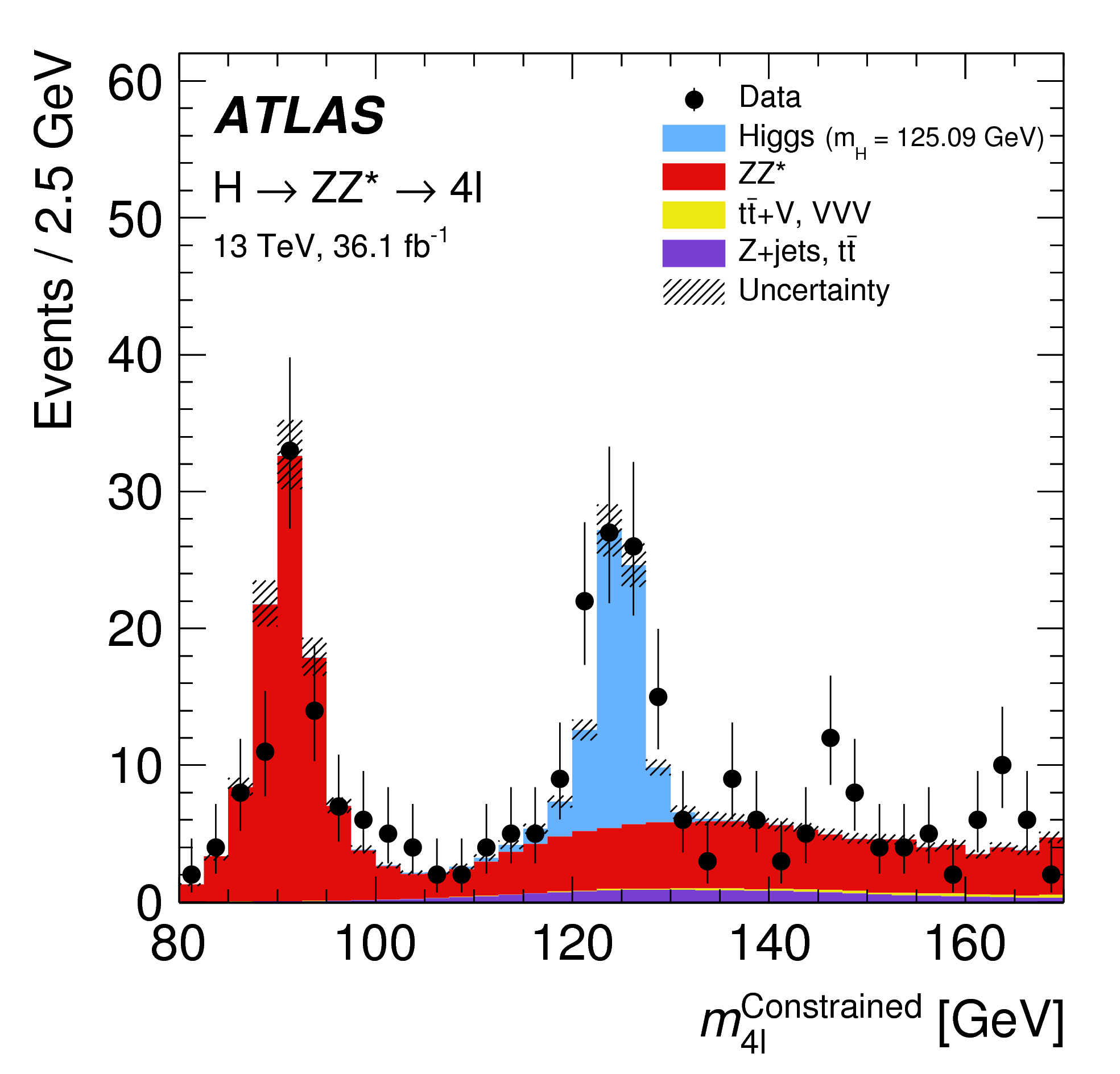}}
\caption{Spectrum of the four lepton mass consistent with a decay $ZZ^*$ as measured by the ATLAS experiment
              \cite{bib-ATLAS-hZZ}. The peak around 91 GeV is the $Z^0$ peak. 
             }
\label{fig:ATLAS-Higgs}        
\end{figure}

In Fig.~\ref{fig:ATLAS-Higgs} the observed
the mass distribution of the four leptons (either electrons or muons) from the $ZZ^*$ decays 
is shown. An enhancement 
around 125 GeV can be seen over an almost flat background. 
How this signal is translated into $\sigma_{h\rightarrow ZZ^*} $, will
be outlined assuming a pure electron signal, although in the analysis both electrons and muons are used.
  
In a first step simulation is applied to find selections to reach a good signal to background ratio
(S/B, cp.~Sect.\ref{sec:DatPhys}).
For the analysis events are selected that contain
four electrons of which at least one has a minimum
momentum $p_T$ in the plane transverse to the beam direction of 20 GeV. 
All leptons should have an angle larger than 160 mrad wrt the beam axis. 
This defines the 'fiducial volume', in which the cross section is determined.
Inspecting Eq.~\ref{eq:Xsec}, one needs for the cross section measurement 
the efficiency and the background, i.e. the number of events that originate from 
other processes but also lead to four electrons.

The efficiency is affected by two major contributions. The first one is how many of the
generated electrons are inside the fiducial volume. This is in a first step given by the matrix element,
however, detector effects like the energy scale and resolution of electrons 
(see Sect.~\ref{sec:E-depo}) have to be taken into account. The second is the
detection efficiency of the electrons. Since the matrix element is known and the detector
effects have been validated, the detection efficiency for $\sigma_{h\rightarrow XX }$
is obtained from Monte Carlo simulation by convoluting the physics model with the 
adjusted detector performance.   

The dominant background source is continuum $Z^0Z^*$ production, i.e. without an intermittent Higgs boson
(see red area in Fig.~\ref{fig:ATLAS-Higgs}), which tightly constrained outside the signal 
region around 125 GeV.  Minor backgrounds are for e.g. due to top - pair production.
These are estimated by a data driven method similar to the one that will be discussed in
the next section.

Relative to the theoretical expectation the measured cross section for a SM Higgs is
 
\begin{equation}
\left( \frac{\sigma(h\rightarrow 4l)_{\mathrm{data}} }{\sigma(h\rightarrow 4l)_{\mathrm{SM}} } \right)_{\mathrm{fid} } \ = \
1.28 \pm 0.18 \pm 0.07 \pm 0.07 
\end{equation}

where the uncertainties are statistical, experimental systematic and theoretical, the latter including pdfs.
The experimental systematic uncertainty corresponds to the uncertainty of the validation from simulation.

\subsection{Search for a Stop Quark}
\label{sect:stop}

A large fraction of the analyses at the LHC tries to find BSM signals. 
In this section it will be discussed, how simulation is applied in the  
search for supersymmetry, the most popular BSM model. 
The focus will be on methods to validate simulation of background processes.

More specifically, 
the search for the pair production of the supersymmetric partner of the top quark, 
denoted as stop ($\tilde{t}$), in the decay mode

\begin{equation}
pp \ \rightarrow \ \tilde{t} \bar{\tilde{t} } \ \rightarrow \ t \bar{t} \chi^0 \chi^0
\end{equation}

will be considered, where $\chi^0 $ is a particle that leaves no trace in any detector - and therefore is
also a dark matter candidate. The production and decay properties of this process are rather precisely
predicted such that the distribution of the outgoing partons can be reliably simulated.
Here we assume a stop quark of high mass (of 1 TeV) and a massless $\chi^0$,
following~\cite{bib:stop}.

The experimental signatures of this process are a visible pair of SM top quarks 
($t \bar{t}$), but in association with a very high imbalance of the 
detectable momentum in the plane transverse to the beam axis, caused by the two invisible
$\chi ^0$s. This is denoted as 'missing transverse energy', $E_{\mathrm{T,miss} }$. 
One major background is due to two neutrinos from decays of top quarks, 

\begin{equation}\label{eq:ttnunu}
pp \ \rightarrow \ t \bar{t}  \ \rightarrow \nu \bar{\nu } + X
\end{equation}

where $X$ represents all other particles in the event. The validation of this process
will be discussed here.
Simulations are used to suggest a selection that leads to an optimal S/B ratio for stop production. 
This signal region (SR) is defined through six observables, of which the most important are  

\begin{itemize}
\item $m_T \ > $ 160 GeV, the mass of the lepton $p_T$ and $E_{\mathrm{T,miss} }$ system.
 \item $\mathrm{am}_{T2} \ > $ 175 GeV, a measure in how far the observed jets and leptons agree with coming 
         from two stop quarks of mass 1 TeV\footnote{The exact definition is
         \begin{equation}
         \mathrm{am}_{T2} \ = \ \min_{\mathbf{q}_{Ta}+\mathbf{q}_{Tb} \ = \ E_{\mathrm{T,miss} } } 
                    \left[max(m_{Ta},m_{Tb} )\right]
         \end{equation}
         I.e. the mimimum parent mass consistent with the observed kinematic distributions assuming input
         masses $m_{Ta}$ and $,m_{Tb}$ and certain mass combinations.
}.
\end{itemize}

The retained events are in kinematic regions ,
which have not yet been probed and are prone to possible misrepresentations of both 
Standard Model physics and detector modeling.
Using the formalisms of Sect.~\ref{sect:factorization},  
one can separate the whole distribution for simplicity 
into regions that have, respectively have not, been
probed, i.e. $z < z_{\mathrm{cut} }$ and $z > z_{\mathrm{cut} } = (z_{\mathrm{cut} }+\Delta)$. 
The $z_{\mathrm{cut} }$ may be identified with the selection requirement used to isolate stop pair - events
and the region $(z_{\mathrm{cut} }+\Delta) $ with the SR.   
In this region simulation of the Standard Model distributions is given by

\begin{equation}\label{eq:DeltaM}
P'(z_{\mathrm{cut} }+\Delta) \  =  \  \left[ {\cal{M} }  + \Delta{\cal{M} } \right] \cdot (B+U)(z_{\mathrm{cut}} +\Delta)  
\end{equation}

$\Delta{\cal{M} }$ represents the migration that has not been tested before,
whereas $B(z_{\mathrm{cut}}+\Delta) $ the Standard Model distribution in the not yet tested region.

It follows from Eq.~\ref{eq:DeltaM} that the contribution to $P(z>z_{\mathrm{cut}} )$ is due 
to two sources: the Standard Model contribution $B(z>z_{\mathrm{cut} })$ and the migration of events out of
$B(z<z_{\mathrm{cut} })$ (and of course both together). The simulation of 
the background yield, needs to be validated in the region $(z_{\mathrm{cut} }+\Delta)$. 
This is done in two steps

\begin{figure}[hbt]
\includegraphics[scale=.08]{{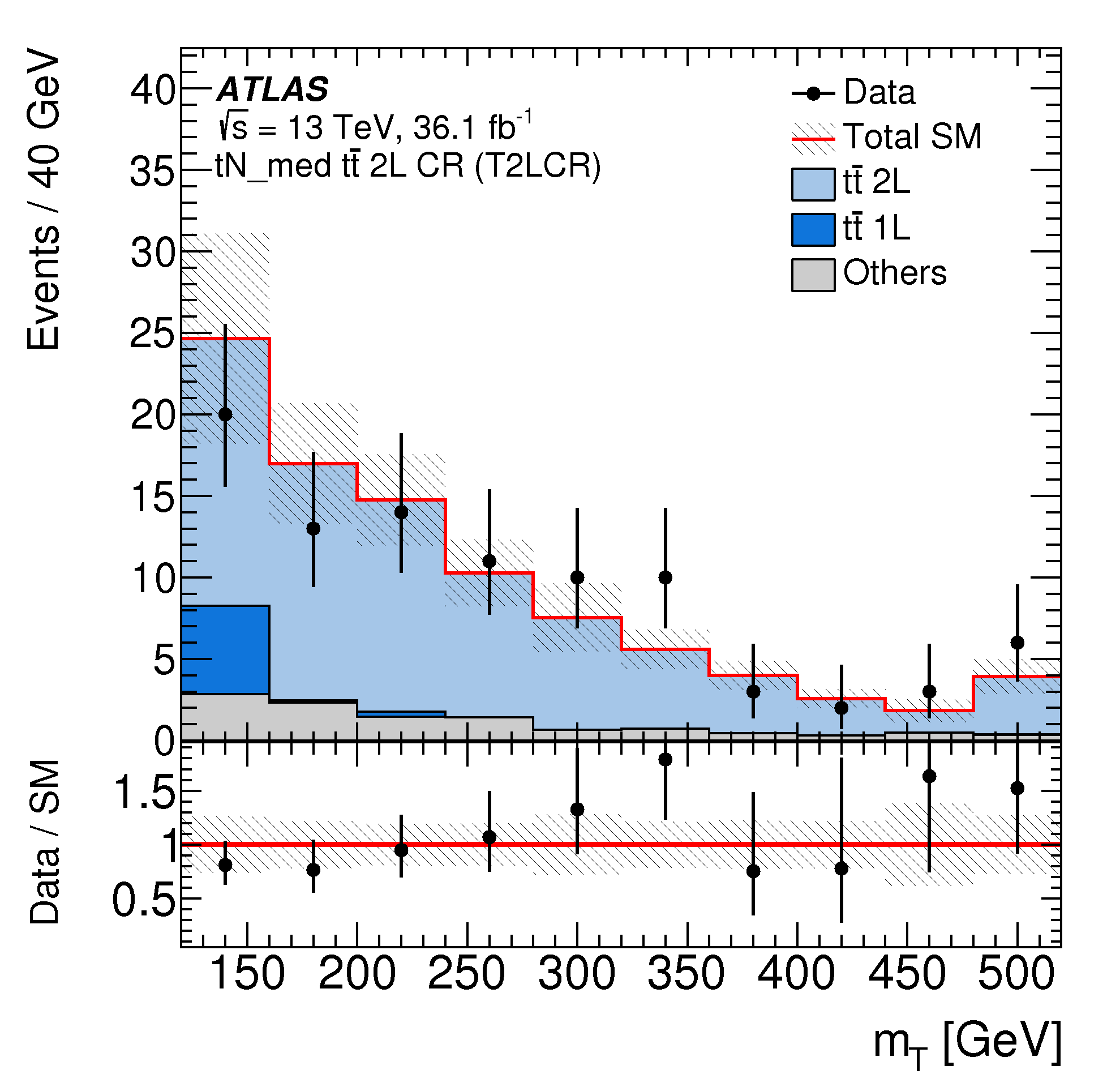}}
\includegraphics[scale=.08]{{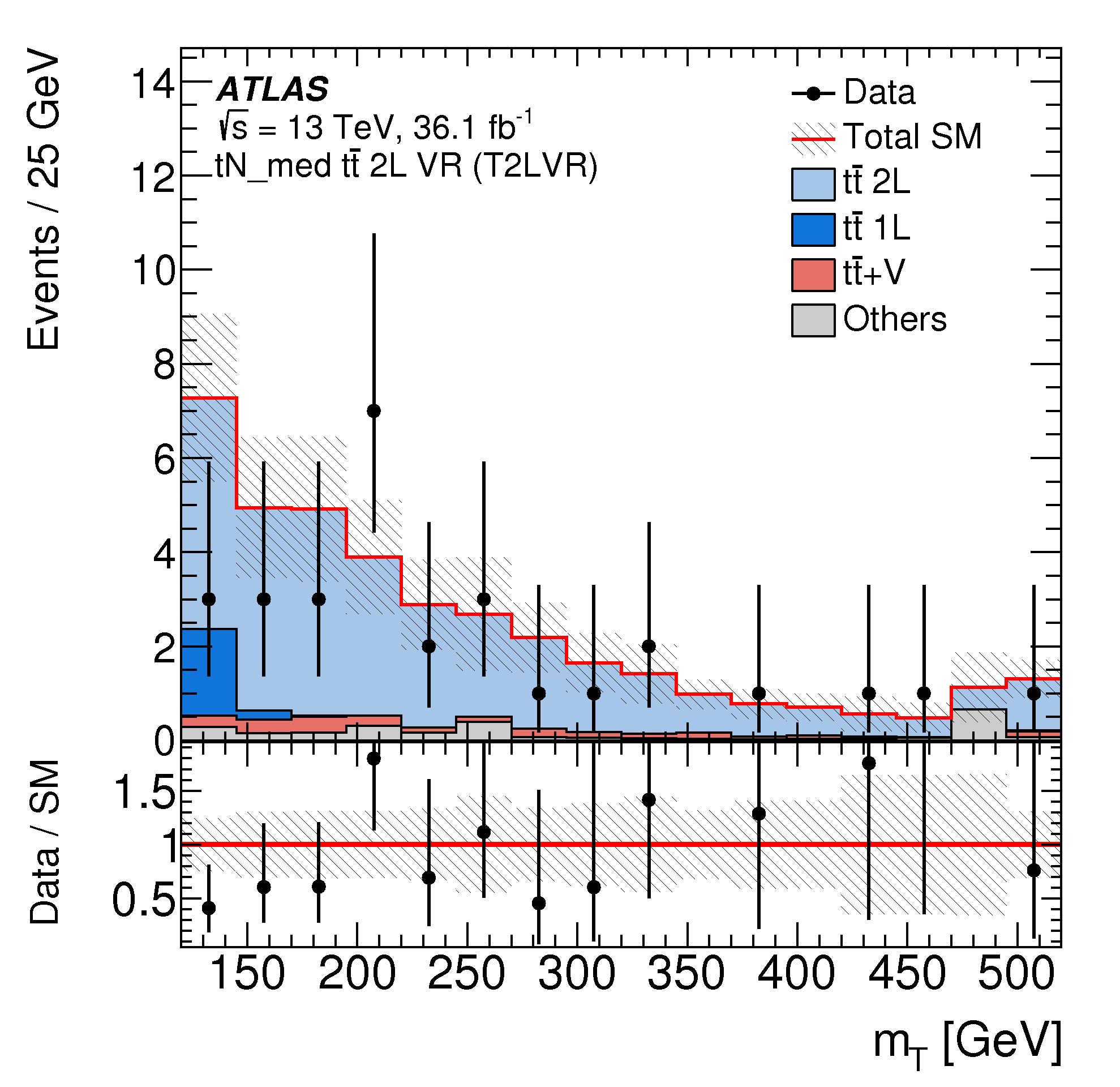}}
\caption{The $m_T$ distribution for the control region (CR, left) and the validation region (VR, right). Shown are
              the expectations from the dominant Standard Model contributions from $t\bar{t}$ production and the data.
              The lower part shows the ratio data/simulated events~\cite{bib:stop}.}
\label{fig:CRVRSR}        
\end{figure}

\begin{itemize}
\item In a first step, a 'control region' (CR) for the process~\ref{eq:ttnunu}
         is defined, which uses the same observables as the SR, with cut values in general 
         as close as possible to the SR, but inverting the $\mathrm{am}_{T2}$  requirement.
         This enriches $t\bar{t}$ events, makes the SR and CR regions disjunct 
         and the CR void of any signal. 
         The normalized distributions (shapes) are sensitive to detector effects 
         and data and simulation are compared for $z>z_{\mathrm{cut} }$. 
         As an example the $m_T$ distribution is depicted in
         Fig.~\ref{fig:CRVRSR} (left) showing a good agreement. 
         It underlines that the physics distributions and the detector effects are well 
         described also for $m_T >$ 160 GeV, which is sensitive to a potential stop particle.
         Once the shape is confirmed, the simulated cross section for 
         $z>z_{\mathrm{cut} }$ is scaled by\footnote{Note that using this method, 
         other backgrounds, like
         $W$+jets, show a visible discrepancy between simulation and data.}.
         \begin{equation}
         r_{t\bar{t} } \ = \ \left(
         \frac{N_{\mathrm{data}}(t\bar{t} )}{N_{\mathrm{simulation}}(t\bar{t} )} 
         \right)_{\mathrm{CR} }
         \ = \ 1.01 \pm  0.15
         \end{equation}
         and therefore adjusted to the measurement.
                     
\item In a second step, a 'validation region' (VR) is defined where the selection is chosen to be in
         between the CR and the SR. 
         A small, but negligible, fraction of events
         might come from the signal. Using the adjusted cross section for the simulation of the background, 
         it is tested in how far the observed number and shape of events agrees with the expectation.
         Fig.~\ref{fig:CRVRSR} (right) shows that the data can be consistently described.
                       
\end{itemize}

These studies validate the background distributions in the kinematic vicinity of the SR, but not
in the SR itself. After having adjusted simulation to data in the CR and VR, one uses simulation
to extrapolate. However, since these extrapolations depend on details of strong interactions  
(see Sect.~\ref{sec:dressing}), a range of QCD models is used to estimate its additional
uncertainty taking into account the constraints from the CRs and VRs. 

I.e. the background in the new region is estimated with methods which use simulation as guidance,
but in the validation process they are adjusted to agree with the data. This procedure implies a
significant uncertainty such that the search is sensitive only in regions where the S/B is high.
In the stop analysis the expected number of Standard Model  
background events in the SR of 3.8$\pm $1.0, the uncertainty mostly due to the 
modeling uncertainties. The expected signal contribution from a stop would be 6. 
In the data 8 events are observed,
i.e. more than expected from SM sources alone, but consistent with just a statistical fluctuation.
  
\section{Discussion}

Simulation in science and its validation are debated in the philosophical literature from various
perspectives.
In this short section a few points are commented
from the view of particle physics without being able to address them in detail.


Simulation of LHC events is characterized by the remoteness
of the underlying physics and the complexity of the measurement device. As a result
it involves hugely different scales. To say
it strikingly, simulation covers the physics from distances of 10$^{-18}$ meters to
those of several meters in the detector. For the different scales specific models are employed, 
such that instead of one model a chain of models is used in simulation. 
This is the basis of factorization addressed in this article.

Simulations allow one to take into account non - linear effects and stochastic
distributions and thus a more detailed modeling than analytical calculations. 
Simulations are therefore instrumental to improve the precision of the
measurements and their interpretations. 
At least most of the principle techniques of data analysis using simulations are similar to those
that have been invoked before. E.g. it was a standard method in analytical $\chi^2$ minimizations
to vary parameters in a model and see which one fits best. 
Simulations allow one to account for also subtle effects and many parameter variations
with detailed templates. 
While such parameter variations motivate some philosophical literature to consider simulation as
an 'experiment' by itself \cite{bib:EnzSimu} they are no new quality of 
scientific practice\footnote{New potentials through simulations 
may have been opened for using machine learning techniques in data analysis.}.  

The higher precision comes at the price of higher complexity which
naturally reflects itself in the complexity of validation. Required is
the validation of each model in the chain since incorrectness of one of the models 
implies the whole simulation to be incorrect. Since  simulation is
factorizable, the complexity of validation is broken down into the validation of
many 'simple' models. 
To the extent that the factorization is exhaustive and each factor is validated, the 
complete simulation is validated.  

The models applied used have different confirmation statuses. 
Most of these are embedded in well established scientific practices that have grown and
been confirmed over decades. Once such models have attained a very high
level of confirmation, they acquire an almost autonomous status in simulation, 
i.e. their predictions are largely
accepted. Other models are less certain, calling for more detailed scrutiny. 
E.g. simulation of the detector response for electrons is 
significantly better known than the emergence of jets from partons. 
Such models are less trusted and call for specific care in validation.

However, even if models are strongly confirmed, there is a reluctance among particle 
physicists to rely too strongly on these.  
For once all models  assume certain parametrizations and parameter values and it may
depend on special circumstances if these are applicable. E.g. even though the
interactions of an electron in a material are well known, the distribution of the material may
be not. Therefore, in - situ validations or data driven methods are used to a large part.  
As discussed in this article, the individual model predictions at the LHC can be tested 
directly with complementary processes without any circular argumentation and are rather tightly constrained by
the data. This appears to be in contrast to the claim of Morrison (cp.Sect.~\ref{sec:genproc}). 
In all these cases simulation is adjusted to describe the data - 
not vice versa, pointing to very different epistemic statuses of data and simulation.

Validation, however, has to account for the obvious fact that simulations cannot 
describe phenomena in all fine details. 
There is no validation 'per se' but only a validation with some
uncertainty - validation in particle physics has to be quantifiable, 
leading to a systematic uncertainty of the simulation, which is 
a systematic uncertainty of the understanding of the measurement.
At the LHC 
validation of the simulation and assigning a systematic uncertainty is almost synonymous
and most of the work in LHC's data analysis is devoted to estimating these uncertainties. 
For some processes at the LHC the systematic uncertainty is reduced to the 0.1\% level, 
for some they are much higher.

\section{Summary and Conclusion}

Simulation in particle physics is performed in factorizable steps that can be interpreted
as migrating a 'true' distribution of an underlying parton process to the measurable one. 
This factorization allows one to
validate the individual steps using specific measurements in which these steps
are isolated and circularity with the target physics process is avoided.

In the validation process several models of different confirmation status are used. Some
have been found to agree with data in many different experiments in the past, others
are relatively new. This confirmation status affects how these models are applied, 
respectively which uncertainties are assigned. In all cases, particle physicists take pain 
to validate the models and their application using data, trying 
to minimize the reliance of simulation and its validation on models. 
One can distinguish the following methods: 

\begin{itemize}
\item in - situ calibration of detector and physics processes,
\item adopting previous precision measurements as references,
\item deriving models from previous measurements and applying them to LHC data.
\end{itemize}

These different methods rely on data and models at different levels,
also implying different ways to estimate the uncertainties.
 
In conclusion, substantial effort in particle physics is devoted to the validation of 
event simulation. The validation is in all steps significantly constraint by data. Models 
as autonomous entities are only
invoked if they have been strongly confirmed in previous measurements. However,
even in this case, the simulation is checked in an actual experiment.
A quantitative estimate of the validity of the simulation is based on the
factorization of individual contributions. Highly efficient methods have been devised to 
estimate and minimize the uncertainties with a strong effort to constrain those from data -
either directly or reducing model choices. \\ \\

\textbf{Acknowledgement}
I am grateful to Martin King and Michael St\"oltzner and anonymous referees 
for valuable comments. I also profited highly 
from discussions with colleagues from the Research Group 'Epistemology of the LHC'
funded by the DFG under grant FOR 2063.

\end{document}